\documentclass[aps,prl,twocolumn,groupedaddress]{revtex4-1}

\usepackage{enumitem} %for customizable list formatting
\usepackage{graphicx} %for importing pdf figures
\usepackage{multirow} %for table formatting
\usepackage{array} %for the following definitions of columns used in formatting tables
\newcolumntype{L}[1]{>{\raggedright\let\newline\\\arraybackslash\hspace{0pt}}m{#1}} %left-justified column with supplied width
\newcolumntype{C}[1]{>{\centering\let\newline\\\arraybackslash\hspace{0pt}}m{#1}} %center-justified column with supplied width
\newcolumntype{R}[1]{>{\raggedleft\let\newline\\\arraybackslash\hspace{0pt}}m{#1}} %right-justified column with supplied width

\begin{document}

\title{Gender inequities throughout STEM: Women with higher grades drop STEM majors while men persist}

\author{Kyle~M.~Whitcomb}
\affiliation{Department of Physics and Astronomy, University of Pittsburgh, Pittsburgh, PA, 15260}
\author{Chandralekha~Singh}
\affiliation{Department of Physics and Astronomy, University of Pittsburgh, Pittsburgh, PA, 15260}

\date{\today}

\begin{abstract}
	Efforts to promote equity and inclusion using evidence-based approaches are vital to correct long-standing societal inequities that have disadvantaged women and discouraged them from pursuing studies, e.g., in many STEM disciplines.
	We use 10 years of institutional data at a large public university to investigate trends in the majors that men and women declare, drop after declaring, and earn degrees in as well as the GPA of the students who drop or earn a degree.
	We find that the majors with the lowest number of students also have the highest rates of attrition.
	Moreover, we find alarming GPA trends, e.g., women who drop majors on average earn higher grades than men who drop those majors, and in some STEM majors, women who drop the majors were earning comparable grades to men who persist in those majors.
	These quantitative findings call for a better understanding of the reasons students drop a major and for making learning environments equitable and inclusive.
\end{abstract}

\maketitle

\section{Introduction and Theoretical Framework}

Increasingly, Science, Technology, Engineering, and Mathematics (STEM) departments across the US are focusing on using evidence to improve the learning of all students, regardless of their background and making learning environments equitable and inclusive~\cite{johnson2012, johnson2017, metcalf2018, king2016, maltese2011, maltese2017, means2018, borrego2008, borrego2011, borrego2014, henderson2008, dancy2010, henderson2012}.
However, women are still severely underrepresented in many STEM disciplines~\cite{nsc2015, nsf2018}.
In order to understand the successes and shortcomings of the current state of education, the use of institutional data to investigate past and current trends is crucial.
In the past few decades, institutions have been keeping increasingly large digital databases of student records.
We have now reached the point where there are sufficient data available at many institutions for robust statistical analyses using data analytics that can provide invaluable information for transforming learning environments and making them more equitable and inclusive for all students~\cite{baker2014, papamitsiou2014}.
Studies utilizing many years of institutional data can lead to analyses that were previously limited by statistical power.
This is particularly true for studies of performance and persistence in STEM programs that rely on large sample sizes~\cite{ohland2008, lord2009, eris2010, maltese2011, min2011, lord2015, ohland2016, matz2017, witherspoon2019, king2016, safavian2019, maltese2017, means2018}.

In this study, we use 10 year institutional data from a large state-related research university to investigate how patterns of student major-declaration and subsequent degree-earning may differ for men and women.
The theoretical framework for this study has two main foundations: critical theory and expectancy value theory.

Critical theories, e.g., of race and gender, focus on historical sources of inequities within society, that is, societal norms that perpetuate obstacles to the success of certain groups of disadvantaged people~\cite{gopalan2019, crenshaw1995, kellner2003, yosso2005, gutierrez2009, taylor2009, tolbert2018, schenkel2020, metcalf2018}.
Critical theory tells us that the dominant group in a society perpetuates these norms, which are born out of their interests, and pushes back against support systems that seek to subvert these norms~\cite{crenshaw1995, kellner2003, yosso2005}.
In our case, critical gender theory provides a historical perspective on the much-studied gender inequities in STEM.

Much important work has been done that relates to critical theory of gender in STEM education~\cite{johnson2012, johnson2017, bang2010, estrada2018, ong2018, tolbert2018, schenkel2020, lord2009, seron2016, brawner2012, brawner2015, metcalf2018, ganley2018}.
One mechanism by which historical societal stereotypes and biases about gender can influence student choice of major is proposed by Leslie \textit{et al.}, who showed that disciplines with a higher attribution of ``brilliance'' also have a lower representation of women~\cite{leslie2015} due to pervasive stereotypes about men being ``brilliant'' in those disciplines.
These brilliance-attributions affect all levels of STEM education, starting with early childhood where girls have already acquired these notions that girls are not as brilliant as boys~\cite{bian2017, bian2017thesis}, which can later influence their interest in pursuing certain STEM disciplines~\cite{bian2018messages}, and even affect how likely they are to be referred for employment in these disciplines in professional contexts~\cite{bian2018evidence}.

Expectancy value theory (EVT) is another framework that is central to our investigation and states that a student's persistence and engagement in a discipline are related to student's expectancy about their success as well as how the student values the task~\cite{eccles1984, eccles1990, eccles1994}.
In an academic context, ``expectancy,'' which refers to the individual's beliefs about their success in the discipline, is closely related to Bandura's construct of self-efficacy, defined as one's belief in one's capability to succeed at a particular task or subject~\cite{bandura1991, bandura1994, bandura1997, bandura1999, bandura2001, bandura2005, eccles1984, eccles1990, eccles1994}.

There are four main factors that influence students' expectancy or self-efficacy, namely vicarious experiences (e.g., instructors or peers as role models), social persuasion (e.g., explicit mentoring, guidance, and support), level of anxiety~\cite{bandura1991, bandura1994, bandura1997, bandura1999, bandura2001, bandura2005}, and performance feedback (e.g., via grades on assessment tasks).
Women generally have lower self-efficacy than men in many STEM disciplines because these four factors negatively influence them~\cite{johnson2012, johnson2017, bang2010, estrada2018, ong2018, tolbert2018, schenkel2020}.
For example, in many STEM fields women are underrepresented in their classrooms, and less likely to have a female role model among the faculty~\cite{nsc2015, nsf2018}.
Further, the stereotypes surrounding women in many STEM disciplines can affect how they are treated by mentors, even if such an effect is subconscious~\cite{astin1993, johnson2012, felder1995, felder1998, cheryan2017, bianchini2002, britner2006, hilts2018}.
Moreover, women are susceptible to stress and anxiety from stereotype threat (i.e., the fear of confirming stereotypes about women in many STEM disciplines) which is not experienced by their male peers~\cite{astin1993, johnson2012, felder1995, felder1998, cheryan2017, bianchini2002, britner2006, hilts2018}.
This stress and anxiety can rob them of their cognitive resources, especially during high-stakes assessments such as exams.

Expectancy can influence grades earned as well as the likelihood to persist in a program~\cite{bandura1991, bandura1994, bandura1997, bandura1999, bandura2001, bandura2005}.
Stereotype threats that women in many STEM disciplines experience can increase anxiety in learning and test-taking situations and lead to deteriorated performance.
Since anxiety can increase when performance deteriorates, these factors working against women in STEM can force them into a feedback loop and hinder their performance further, which can further lower their self-efficacy and can continue to affect future performance~\cite{bandura1991, bandura1994, bandura1997, bandura1999, bandura2001, bandura2005}.

In EVT, value is typically defined as having four facets: intrinsic value (i.e., interest in the task), attainment value (i.e., the importance of the task for the student's identity), utility value (i.e., the value of the task for future goals such as career), and cost (i.e., opportunity cost or psychological effects such as stress and anxiety)~\cite{eccles1984, eccles1990, eccles1994}.
In the context of women's enrollment and persistence in many STEM disciplines, the societal stereotypes can influence all facets of the students' value of these STEM disciplines.
Intrinsic value can be informed by societal stereotypes and brilliance-attributions of the STEM disciplines, and attainment and utility values can be further tempered by these stereotypes.
Utility value is an important facet of student education in STEM, since a degree in a STEM field provides many job opportunities for graduating students.
In addition, the psychological cost of majoring in these disciplines can be inflated by the stereotype threat.
All of these effects can conspire to suppress the likelihood of women choosing and/or persisting in various STEM disciplines.

In order to measure the long-term effects of these systemic disadvantages, we investigate the differences in attrition rates and choices of major of men and women over the course of their studies at one large public research university using 10 years of institutional data.
Since these disadvantages to students can be context-dependent, we will consider the attrition rates in many different STEM majors and non-STEM majors in order to understand the trends in each discipline.

\subsection{Research Questions}

Our research questions regarding the relationships between gender, degree attainment, attrition, performance and persistence pertaining to a college degree over a 10 year period are as follows.

\begin{enumerate}[label={\bfseries RQ\arabic*.}, ref={\bfseries RQ\arabic*}, itemsep=1pt]
	\item \label{rq_declare} How many students major in each discipline? How many men and women major in each discipline?
	\item \label{rq_drop} Do rates of attrition from the various majors differ? Do rates of attrition from the various majors differ for men and women?
	\item \label{rq_droppers} Among those students who drop a given major, what degree, if any, do those students earn? How do these trends differ for men and women?
	\item \label{rq_degree} What fraction of declared majors ultimately earn a degree in that major in each STEM subject area? How do these trends differ for men and women?
	\item \label{rq_gpa} What are the GPA trends over time among students who earn a degree in a given major and those who drop that major? How do these trends differ for men and women?
\end{enumerate}

\section{Methodology}

\subsection{Sample}

Using the Carnegie classification system, the university at which this study was conducted is a public, high-research doctoral university, with balanced arts and sciences and professional schools, and a large, primarily residential undergraduate population that is full-time and reasonably selective with low transfer-in from other institutions~\cite{carnegie}.

The university provided for analysis the de-identified institutional data records of students with International Review Board approval.
In this study, we examined these records for $N = 18,319$ undergraduate students enrolled in two schools within the university: the School of Engineering and the School of Arts and Sciences.
This sample of students includes all of those from ten cohorts who met several selection criteria, namely that the students had first enrolled at the university in a Fall semester from Fall 2005 to Fall 2014, inclusive, had provided the university with a self-reported gender, and the students had either graduated and earned a degree, or had not attended the university for at least a year as of Spring 2019.
This sample of students is 49.9\% female and had the following race/ethnicities: 77.7\% White, 11.1\% Asian, 6.8\% Black, 2.5\% Hispanic, and 2.0\% other or multiracial.

\subsection{Measures}

\subsubsection{Gender}

In this study, we focus on gender differences in student trajectories as they progress towards degrees.
We acknowledge that gender is not a binary construct, however in self-reporting their gender to the university students were given the options of ``male'' or ``female'' and so those are the two self-reported genders that we are able to analyze.
The student responses to this question were included in the institutional data provided by the university.
Very few students opted not to provide a gender, and so were not considered in this study.
We used the answers of those students who chose either ``male'' (``M'') or ``female'' (``F'') to group students in order to calculate summary statistics on the measures described in this section.

\subsubsection{Academic Performance}

Measures of student academic performance were also included in the provided data.
High school GPA was provided by the university on a weighted scale from 0-5 that includes adjustments to the standard 0-4 scale for Advanced Placement and International Baccalaureate courses.
The data also include the grade points earned by students in each course taken at the university.
Grade points are on a 0-4 scale with $\text{A}=4$, $\text{B}=3$, $\text{C}=2$, $\text{D}=1$, $\text{F}=0$, where the suffixes ``$+$'' and ``$-$'' add or subtract, respectively, $0.25$ grade points (e.g., $\text{B}-=2.75$), with the exception of $\text{A}+$ which is reported as the maximum 4 grade points.
The courses were categorized as either STEM or non-STEM courses, with STEM courses being those courses taken from any of the following departments: biological sciences, chemistry, computer science, economics, any engineering department, geology and environmental science, mathematics, neuroscience, physics and astronomy, and statistics.
We note that for the purposes of this paper, ``STEM'' does not include the social sciences other than economics, which has been included due to its mathematics-intensive content.

\subsubsection{Declared Major and Degree Earned}

For each student, the data include their declared major(s) in each semester as well as the major(s) in which they earned a degree, if any.
The data were transformed into a set of binary flags for each semester, one flag for each possible STEM major as well as psychology and a general non-STEM category for all other majors.
A similar set of flags was created for the degrees earned by students.
From these flags, we tabulated a number of major-specific measures in each semester, including
\begin{itemize}
	\item current number of declared majors,
	\item number of newly declared majors from the previous semester,
	\item number of dropped majors from the previous semester,
	\item number of retained majors from the previous semester.
\end{itemize}
The total number of unique students that ever declared or dropped a major were also computed.
The subset of students that dropped each major were further investigated and the major in which they ultimately earned a degree, if any, was determined.

Throughout this paper we group the STEM majors into two clusters: chemistry, computer science, engineering, mathematics, and physics and astronomy; and biological sciences, economics, geology and environmental science, neuroscience, and statistics.
When ordering majors (i.e., in figures and tables), the majors will be presented in the order they are listed in the previous sentence (first by group, then alphabetically within each group), followed by non-STEM and psychology.
Further, we group the final three STEM majors (geology and environmental science, neuroscience, and statistics) into a category labeled ``Other STEM'' for figures and tables.
Similarly, ``engineering'' groups together all engineering majors for departments in the School of Engineering at the studied university.
These majors include chemical, computer, civil, electrical, environmental, industrial, and mechanical engineering as well as bioengineering and materials science.

Finally, we will make use of shortened labels for the majors in figures and tables.
These shortened labels are defined in Table~\ref{table_major_labels}.

\begin{table}
	\begin{tabular}{l l}
		Major & Short Label \\
		\hline
		Chemistry				& Chem \\
		Computer Science		& CS \\
		Engineering				& Engr \\
		Mathematics				& Math \\
		Physics and Astronomy	& Phys \\
		Biological Sciences		& Bio \\
		Economics				& Econ \\
		Geology and				& \multirow{2}{*}{Other STEM} \\
		Environmental Science	& \\
		Neuroscience			& Other STEM \\
		Statistics				& Other STEM \\
		Non-STEM				& Non-STEM \\
		Psychology				& Psych
	\end{tabular}
	
	\caption{\label{table_major_labels}
		A list of the majors considered in this study and the shortened labels used to refer to those majors in tables and figures.
	}
\end{table}

\subsubsection{Year of Study}

Finally, the year in which the students took each course was calculated from the students' starting term and the term in which the course was taken.
Since the sample only includes students who started in fall semesters, each ``year'' contains courses taken in the fall and subsequent spring semesters, with courses taken over the summer omitted from this analysis.
For example, if a student first enrolled in Fall 2007, then their ``first year'' occurred during Fall 2007 and Spring 2008, their ``second year'' during Fall 2008 and Spring 2009, and so on in that fashion.
If a student is missing both a fall and spring semester during a given year but subsequently returns to the university, the numbering of those post-hiatus years is reduced accordingly.
If instead a student is only missing one semester during a given year, no corrections are made to the year numbering.

\subsection{Analysis}

For each student, we calculate their grade point average (GPA) across courses taken in each year of study from their first to sixth years.
In addition, we calculate the student's STEM GPA in each year, that is, the GPA in STEM courses alone.
The mean GPA as well as the standard error of the mean is computed for various groupings of students~\cite{freedman2007}.

Further, proportions of students in various groups (i.e., grouped by major and/or gender) are calculated along with the standard error of a proportion~\cite{freedman2007}.
In particular, the proportions we report are
\begin{itemize}
	\item the proportion of students in each major that are men or women,
	\item the proportion of men and women, respectively, that declare each subject as a major,
	\item the proportion of declared majors that drop the major,
	\item the proportion of those who drop each major that earn a degree in another major, and
	\item the proportion of all declared majors that ultimately earn a degree in that major.
\end{itemize}

All analyses were conducted using R~\cite{rcran}, making use of the package \texttt{tidyverse}~\cite{tidyverse} for data manipulation and plotting.

\section{Results}

\subsection{Major Declaration Patterns}
%RQ1: How many students major in each discipline? How many men and women major in each discipline?

\begin{figure*}
    \centering
    
	\includegraphics[width=0.95\textwidth]{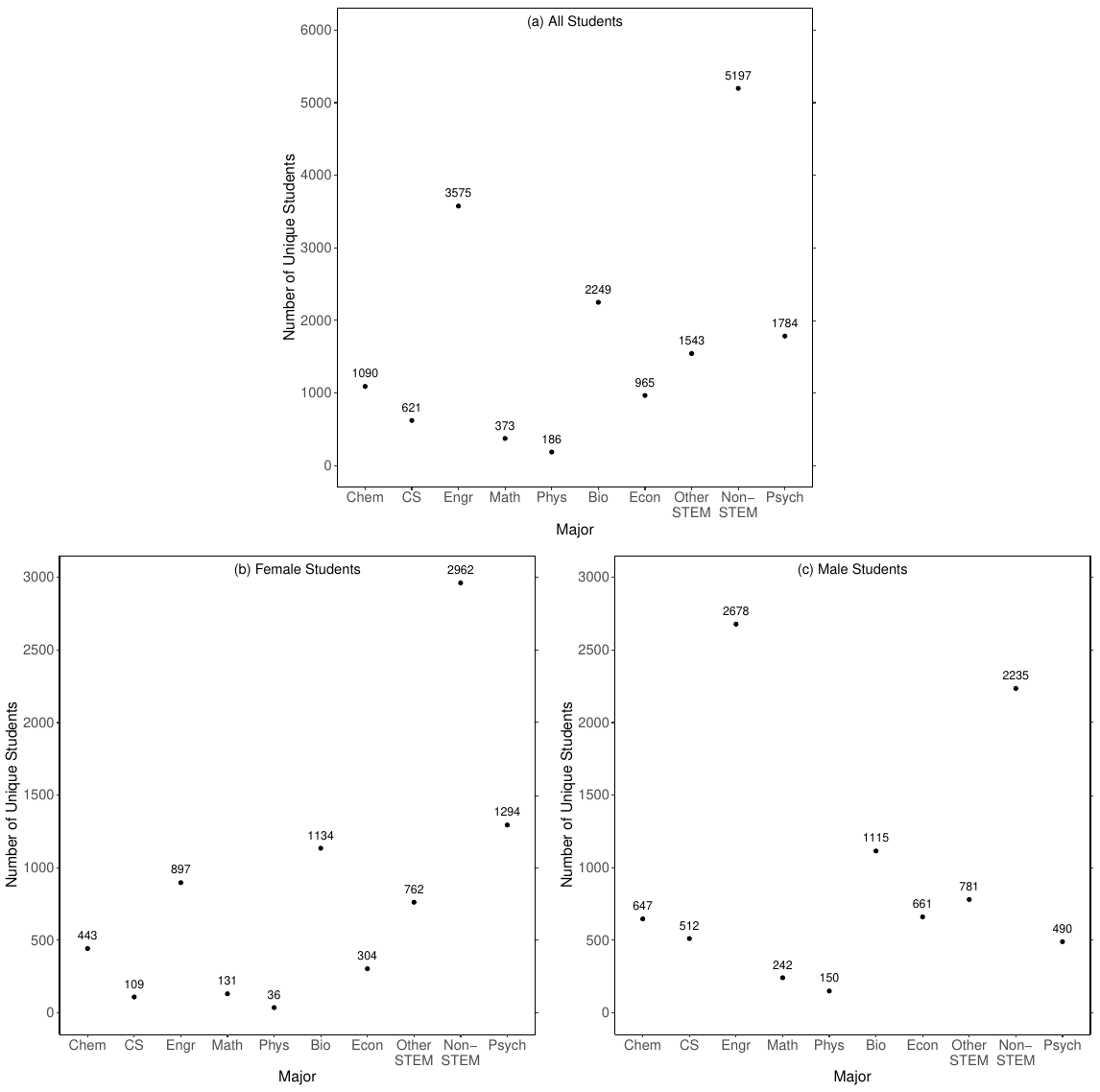}
	
	\caption{\label{figure_n_unique}
	For each major on the horizontal axis, the number of unique students in the sample that ever declared that major is listed.
	Since students may change majors or declare multiple majors, some students may contribute to the counts of more than one major.
	These counts are calculated separately for (a) all students, (b) female students, and (c) male students.
	Note that the scale of the vertical axes differs in (a) compared to (b) and (c).
	}
\end{figure*}

There are many angles with which we can approach \ref{rq_declare} and investigate patterns of student major declaration.
First, Fig.~\ref{figure_n_unique} shows the number of students that ever declared each major.
This is done both overall (Fig.~\ref{figure_n_unique}a) and for female students (Fig.~\ref{figure_n_unique}b) and male students (Fig.~\ref{figure_n_unique}c) separately.
These results provide an important context for the upcoming analyses that may be partially explained by the number of students in each major.

\begin{figure*}
    \centering
    
	\includegraphics[width=0.95\textwidth]{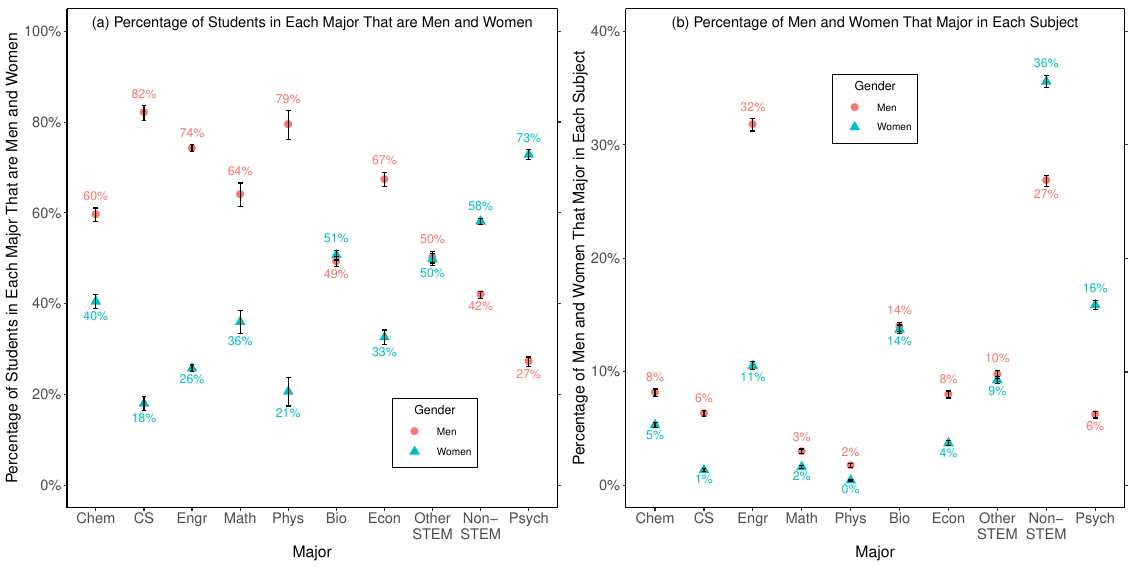}
	
	\caption{\label{figure_percentages}
	In (a), the percentages of students in each major that are men or women are calculated (i.e., the percentages in each column will sum to roughly 100\%).
	In (b), the percentages of men and women that major in each subject are calculated (i.e., the percentages for each gender group will sum to roughly 100\% in this case).
	Discrepancies in the sum of percentages may occur due to rounding the listed percentages to the nearest integer as well as, in (b), students declaring multiple majors.
	}
\end{figure*}

Figures~\ref{figure_n_unique}b and \ref{figure_n_unique}c begin to hint at gender differences in enrollment patterns, such as a higher proportion of women majoring in non-STEM than men, or a higher proportion of men majoring in engineering than women.
These gender patterns are explored further in Fig.~\ref{figure_percentages} by standardizing the scales in two ways.
In Fig.~\ref{figure_percentages}a, we consider the populations of each major separately and calculate the percentage of that population that are men or women.
This provides insight into what these students might be seeing in the classes for their major.
For instance, in the biological sciences there is a roughly even split, so students in biology classes for biology majors might see a classroom that is equally representative of men and women.
On the other end of the spectrum, around 80\% of both computer science and physics and astronomy majors are men.

Another way to represent the population of these majors is to consider what percentage of all men or women choose each major, as seen in Fig.~\ref{figure_percentages}b.
While the main signal of this plot mimics that of Figs.~\ref{figure_n_unique}b and \ref{figure_n_unique}c, we can now see the differences noted earlier more clearly.
In particular, the clearest differences in this view (Fig.~\ref{figure_percentages}b) are in engineering (32\% of men and 11\% of women declare an engineering major), non-STEM (27\% of men and 36\% of women declare a non-STEM major), and psychology (6\% of men and 16\% of women declare a psychology major).

\begin{figure*}
    \centering
    
	\includegraphics[width=0.95\textwidth]{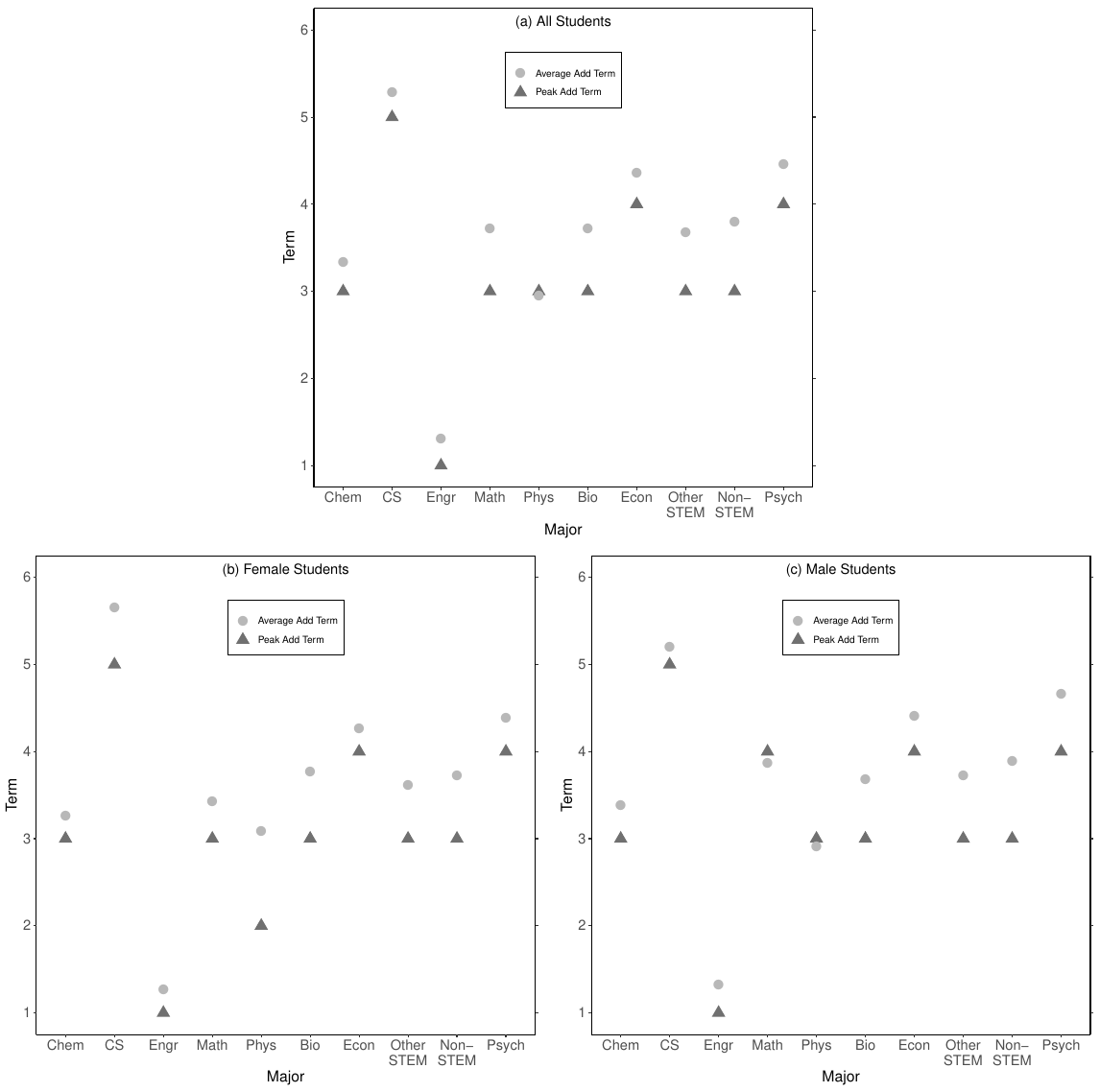}
	
	\caption{\label{figure_peak_avg_term}
	For each major, the term with the peak number of students adding the major in that term is plotted (triangles) as well as the average term in which students add that major (circles).
	This is done separately for (a) all students, (b) female students, and (c) male students.
	}
\end{figure*}

Finally, another piece of information about enrollment patterns that is missing from Figs.~\ref{figure_n_unique} and \ref{figure_percentages} is when these students declare each major.
Figure~\ref{figure_peak_avg_term} shows, for each major, the average term in which students added the major as well as the peak term (that is, the term with the highest number of new students adding the major).
As with Fig.~\ref{figure_n_unique}, this is done separately for all students (Fig.~\ref{figure_peak_avg_term}a), female students (Fig.~\ref{figure_peak_avg_term}b), and male students (Fig.~\ref{figure_peak_avg_term}c).

For the majority of majors in Fig.~\ref{figure_peak_avg_term}a, the peak of students adding the major is in the third term (that is, the start of their second year), with an average of three to four terms.
Two majors, economics and psychology, depart slightly from this general trend, each with a peak in the fourth term and an average between four and five terms.
Two other majors, computer science and engineering, depart more significantly from the general trend in ways that can be explained by their particular implementation at the studied university.

Engineering has a peak in the first term in Fig.~\ref{figure_peak_avg_term}a, with an average only slightly later.
Since all students who enroll in the School of Engineering are considered ``undeclared engineering'' majors (the specific sub-discipline within engineering is not assigned in the first year), the majority of engineering students can be identified in their first term.
Computer science instead has the latest peak term in Fig.~\ref{figure_peak_avg_term}a, namely in the fifth term with a slightly higher average.
This is due to the structure of the computer science program at the studied university, which does not allow students to declare the major until they have completed five of the required courses for the major.
These trends in engineering and computer science are important to keep in mind while considering the results presented later in this paper, since in computer science we are not able to capture attrition that occurs (of students intending to major) during the terms before a student officially declares a major.
Conversely in engineering, we are able to capture almost all attrition in the first year due to the unique enrollment conditions of engineering students, which is not possible for majors within the School of Arts and Sciences (where students can declare their major at any time after the first year).

Turning then to Figs.~\ref{figure_peak_avg_term}b and \ref{figure_peak_avg_term}c, we see almost identical trends as in Fig.~\ref{figure_peak_avg_term}a.
The two exceptions are that the peak declaration of physics majors for women occurs one semester earlier in the second term (Fig.~\ref{figure_peak_avg_term}b).
This is because the overall enrollment in physics is primarily in two semesters, the second and third, which happens to result in different peaks but similar averages for men and women.
Similarly, the peak declaration of mathematics majors for men occurs one semester later in the fourth term (Fig.~\ref{figure_peak_avg_term}c), again since mathematics majors overall are most likely to declare in the third or fourth term.
Apart from these minor differences, these trends in major declaration term between men and women are virtually identical.

A more detailed accounting of the number of students that enroll in each term for each major are reported in Tables~\ref{table_adddrop_1} and \ref{table_adddrop_2} in Appendix A.
Also, summaries of total number of unique students as well as the peak term and number of concurrent students in each major, students adding each major, and students dropping each major are available in Tables~\ref{table_appendix_all_main}, \ref{table_appendix_F_main}, and \ref{table_appendix_M_main} in Appendix B.

\subsection{Attrition Rates}
%RQ2: Do rates of attrition from the various majors differ? Do rates of attrition from the various majors differ for men and women?

\begin{figure*}
    \centering
    
	\includegraphics[width=0.95\textwidth]{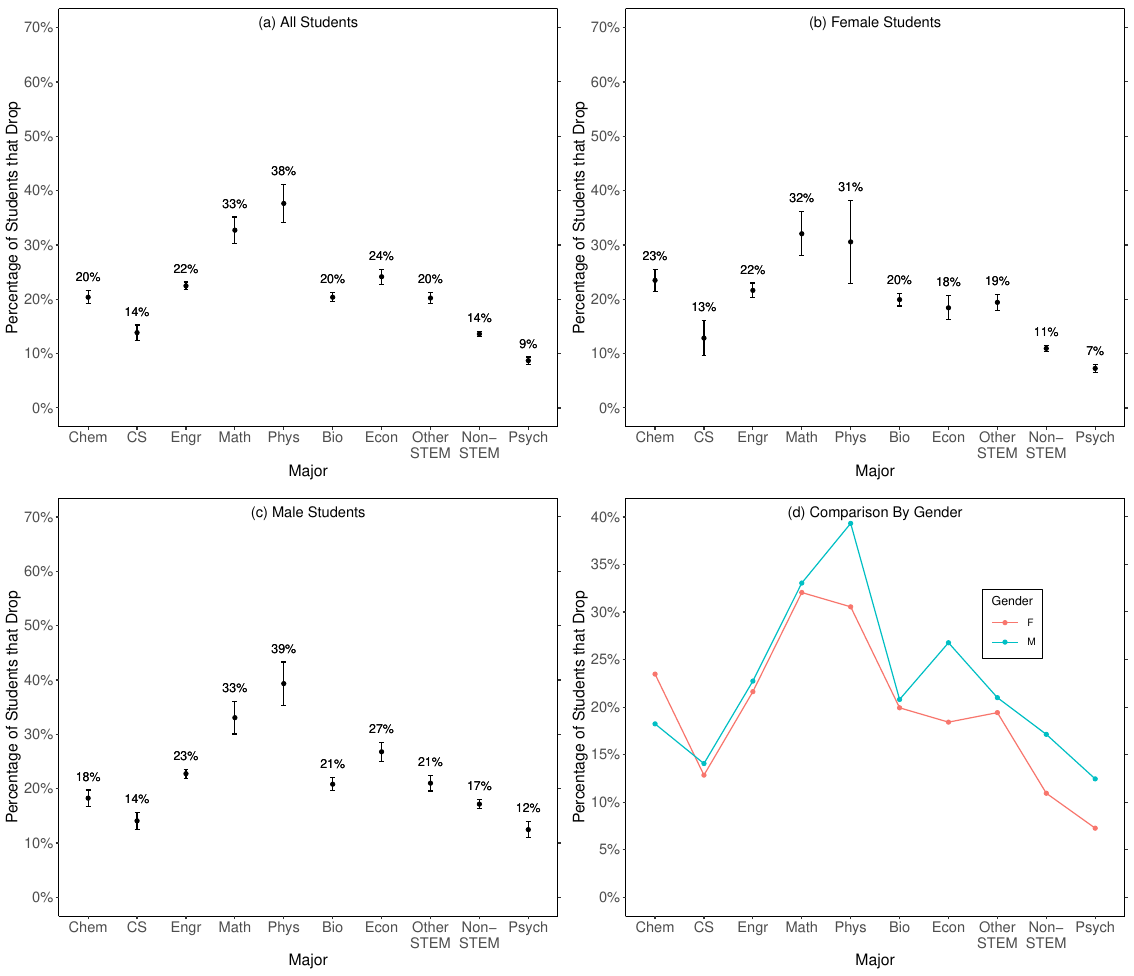}
	
	\caption{\label{figure_drop_plots}
	For each major, the percentage of students who declared the major but subsequently dropped the major is plotted along with its standard error.
	This is done separately for (a) all students, (b) female students, and (c) male students.
	The plots for these men and women are combined into a single plot (d) with the error bars omitted for visibility and ease of comparison, along with lines connecting different points as guides to the eye.
	}
\end{figure*}

In order to answer \ref{rq_drop}, we further considered patterns of attrition rates by gender.
In Fig.~\ref{figure_drop_plots}, we consider the drop rates of different subsets of students (all students, male students, and female students) in each major or group of majors.
In Fig.~\ref{figure_drop_plots}a, we see that computer science, non-STEM, and psychology students are the least likely to drop their major, while physics and mathematics students are the most likely to drop.
We note that the relatively low drop rate of computer science majors could be due to the late declaration of the computer science major seen in Fig.~\ref{figure_peak_avg_term}.
That is, attrition from computer science prior to when students are allowed to declare the major is not accounted for in Fig.~\ref{figure_drop_plots}.

Though the patterns in each subset of students largely mimic the pattern overall in Fig.~\ref{figure_drop_plots}a, there are a few potential exceptions.
Figure~\ref{figure_drop_plots}d shows the drop rates by gender without the error bars, which eases comparison between men and women, however the size of the standard error shown in Figs.~\ref{figure_drop_plots}b and \ref{figure_drop_plots}c indicates that these differences may not be statistically significant.
In particular, in physics women are less likely to drop than men, with 31\% of female physics majors dropping the major (Fig.~\ref{figure_drop_plots}b) and .
Similarly, only 18\% of female economics majors drop that major, while 27\% of male economics majors drop that major.

\subsection{Trajectories of Students After Dropping a Major}
%RQ3: Among those students who drop a given major, what degree, if any, do those students earn? How do these trends differ for men and women?

\begin{figure*}
    \centering
    
	\includegraphics[width=0.95\textwidth]{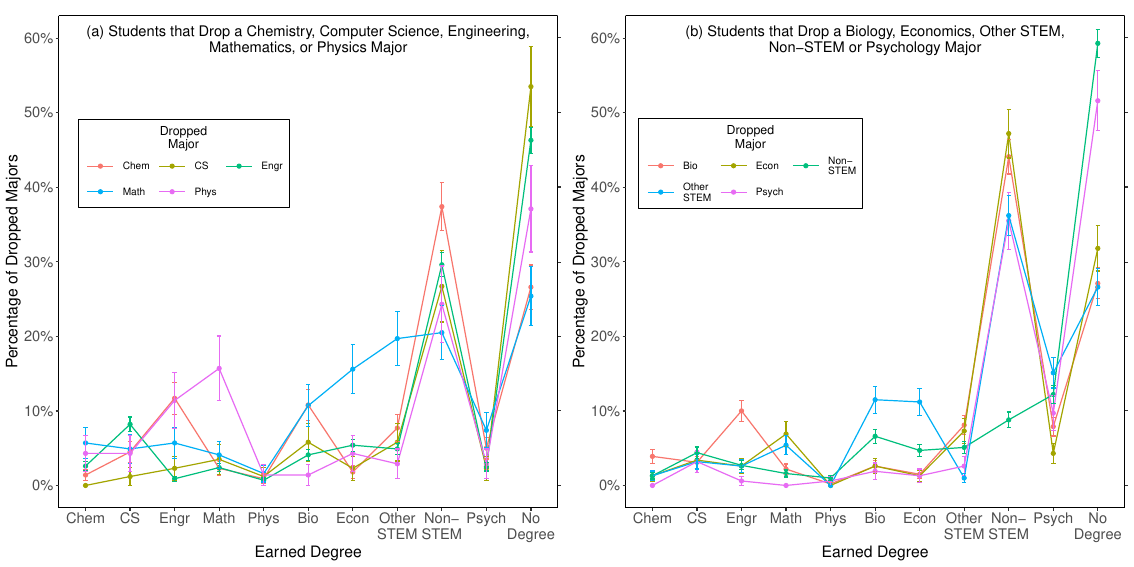}
	
	\caption{\label{figure_drop_plots_degree}
	Among the students that drop each STEM major as well as psychology and non-STEM majors, the fractions of students that go on to earn a degree in other majors, or who do not earn a degree at all, are plotted along with their standard error.
	Dropped majors are grouped into (a) chemistry, computer science, engineering, mathematics, and physics and astronomy majors, and (b) biological science, economics, other STEM, psychology, and non-STEM majors.
	}
\end{figure*}

After discussing how many students drop each major, we answer \ref{rq_droppers} by plotting in Fig.~\ref{figure_drop_plots_degree} where those dropped majors ended up.
In particular, the major indicated in the legends of Fig.~\ref{figure_drop_plots_degree}a and \ref{figure_drop_plots_degree}b shows which major was dropped, while the plot shows the percentage of those who dropped that major and ultimately earned a degree in each of the majors on the horizontal axis, including the case when``no degree'' was earned.
For example in Fig.~\ref{figure_drop_plots_degree}a, we see that among the students that drop the physics major (indicated by the line color in the legend), roughly 15\% of them end up earning a degree in mathematics (by looking at this line's value above ``Math'' on the horizontal axis).
The figure also shows that the two most common destinations for those who drop any major is either no degree or a degree in non-STEM, except for non-STEM majors who are most likely to earn no degree or a degree in psychology.

Apart from that main signal of dropped STEM majors later earning degrees in non-STEM or leaving the university without a degree, we see a few other interesting spikes.
For instance, those who drop a physics major are likely to earn a degree in mathematics (Fig.~\ref{figure_drop_plots_degree}a) and those who drop chemistry and physics (Fig.~\ref{figure_drop_plots_degree}a) as well as biological science (Fig.~\ref{figure_drop_plots_degree}b) are likely to earn engineering degrees.
Further, those who drop from the ``other STEM'' category (geology and environmental science, neuroscience, and statistics) are likely to major in economics and biology (Fig.~\ref{figure_drop_plots_degree}b).
While all students who drop any major are very likely to earn no degree, the percentage of dropped majors in this category exceeds 50\% for computer science (Fig.~\ref{figure_drop_plots_degree}a), non-STEM, and psychology.

\begin{figure*}
    \centering
    
	\includegraphics[width=0.95\textwidth]{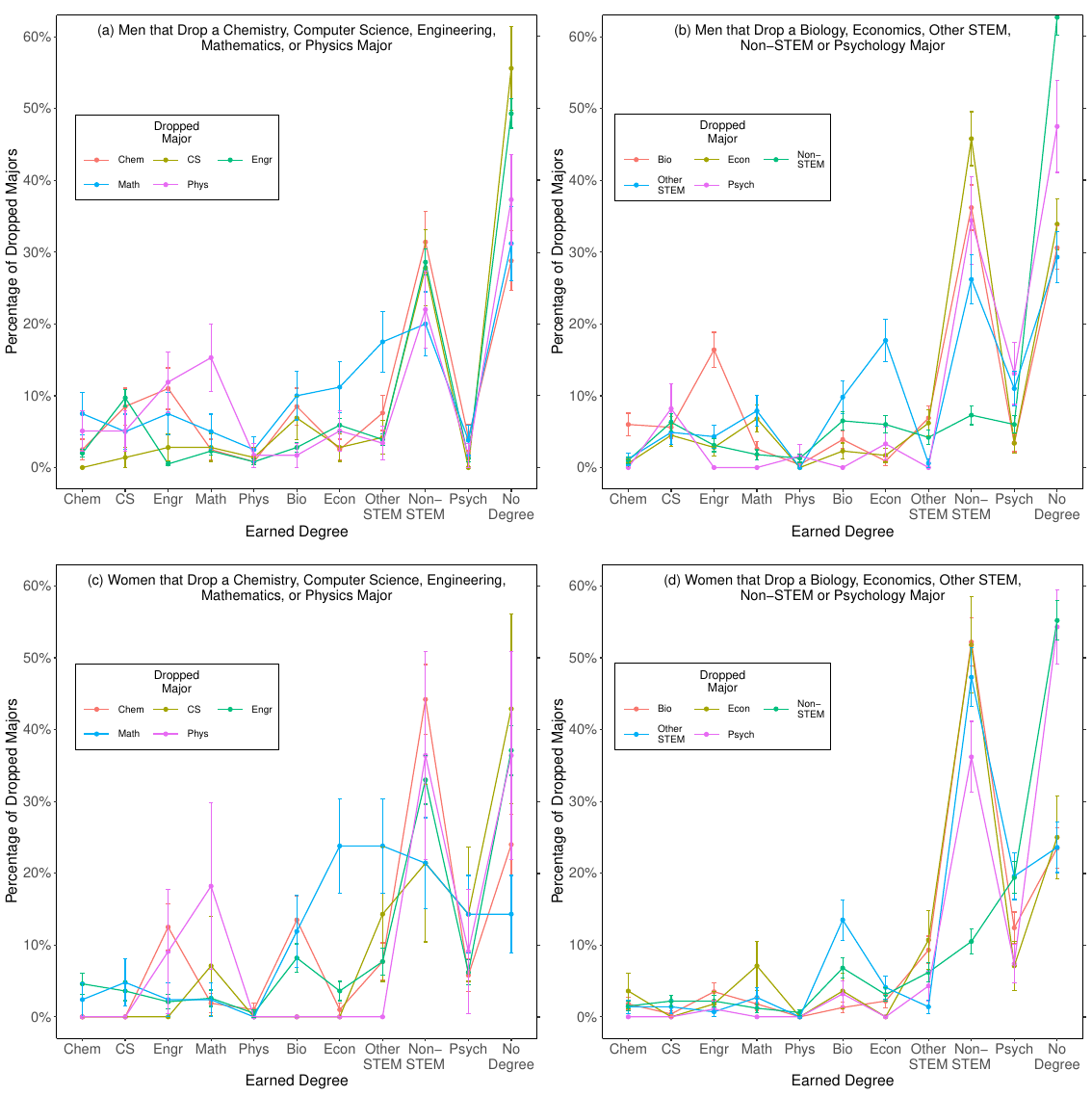}
	
	\caption{\label{figure_drop_plots_degree_by_group}
	Among the men and women that drop each STEM major as well as psychology and other non-SEM majors, the percentages of men and women that go on to earn a degree in other majors, or who do not earn a degree at all, are plotted along with their standard error.
	Dropped majors are grouped into chemistry, computer science, engineering, mathematics, and physics and astronomy majors who are (a) men and (c) women, and biological science, economics, other STEM, and psychology majors who are (b) men and (d) women.
	}
\end{figure*}

In order to further answer \ref{rq_droppers}, Fig.~\ref{figure_drop_plots_degree_by_group} plots these same proportions of degrees earned by students who drop a major separately for men (Figs.~\ref{figure_drop_plots_degree_by_group}a and \ref{figure_drop_plots_degree_by_group}b) and women (Figs.~\ref{figure_drop_plots_degree_by_group}c and \ref{figure_drop_plots_degree_by_group}d).
We see for the most part very similar patterns between men and women, with a few notable differences.
For example, among students who drop a mathematics degree, we see that roughly 23\% of the women eventually earn a degree in economics (Fig.~\ref{figure_drop_plots_degree_by_group}c) compared with roughly 10\% of the men (Fig.~\ref{figure_drop_plots_degree_by_group}a).
We see a similar pattern with the roles reversed among those students who drop a major in the Other STEM category (see Table~\ref{table_major_labels}), with roughly 18\% of the men earning a degree in economics (Fig.~\ref{figure_drop_plots_degree_by_group}b) compared with only 5\% of the women (Fig.~\ref{figure_drop_plots_degree_by_group}d).

A few other examples of gender differences in the trajectories of those who drop a major are that men are more likely than women to earn computer science degrees after dropping a chemistry major (Figs.~\ref{figure_drop_plots_degree_by_group}a and \ref{figure_drop_plots_degree_by_group}c), and similarly men are more likely than women to earn engineering degrees after dropping a biological science major (Figs.~\ref{figure_drop_plots_degree_by_group}b and \ref{figure_drop_plots_degree_by_group}d).
Finally, we note that across all of Fig.~\ref{figure_drop_plots_degree_by_group} in every major except psychology, the women who drop that major are more likely than the men to earn a degree in another major rather than leaving the university (that is, the women have a lower rate of earning ``No Degree'').

A more detailed accounting of the degrees earned by students who drop each major is provided in Tables~\ref{table_appendix_all_supp}, \ref{table_appendix_F_supp}, and \ref{table_appendix_M_supp} in Appendix C.

\subsection{Degree-Earning Rates}
%RQ4: What fraction of declared majors ultimately earn a degree in that major in each STEM subject area? How do these trends differ for men and women?

\begin{figure*}
    \centering
    
	\includegraphics[width=0.95\linewidth]{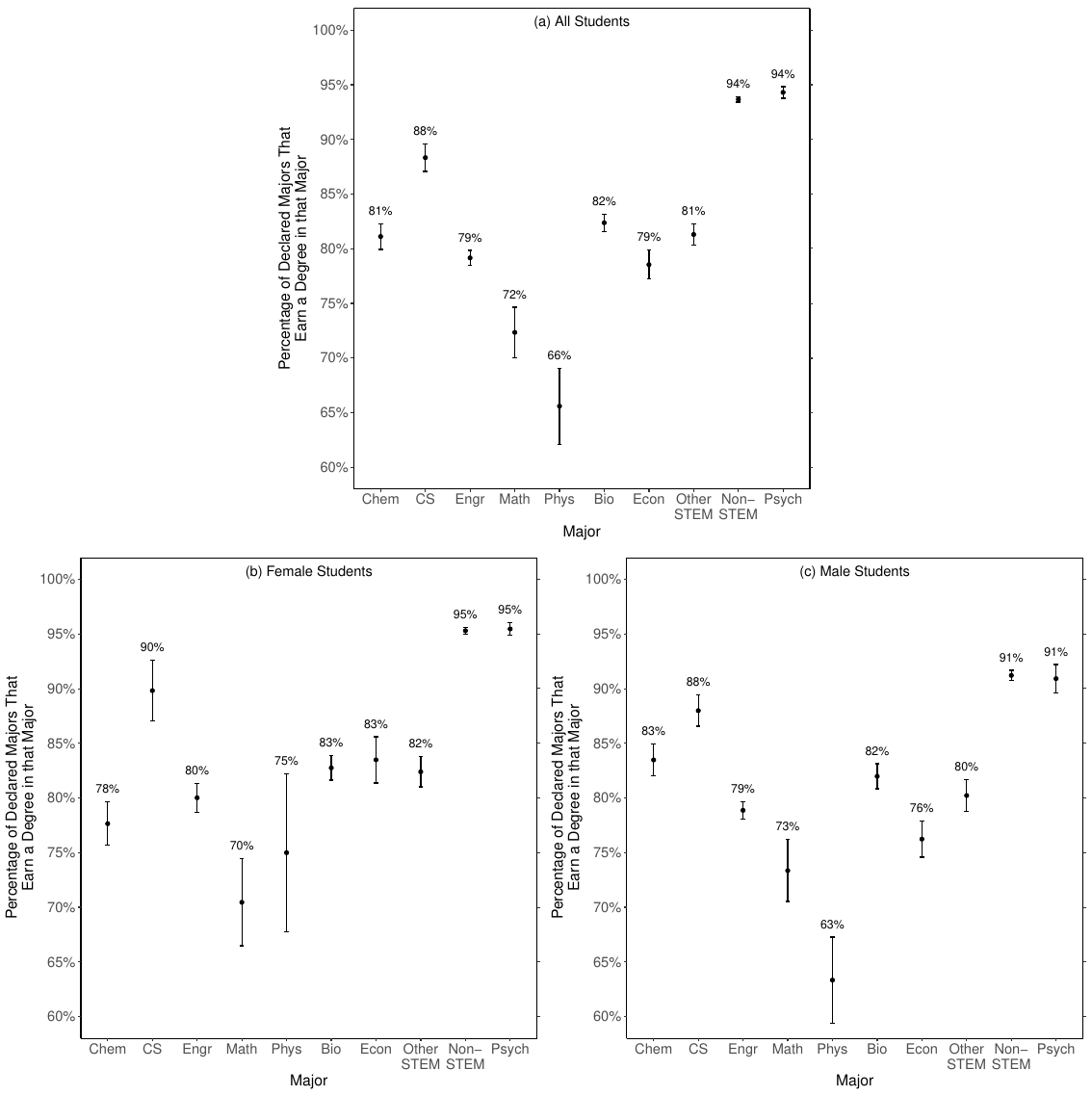}
	
	\caption{\label{figure_degree_rates}
	For each major listed on the horizontal axis, the percentages of (a) all students, (b) female students, and (c) male students who declare that major and then earn a degree in that major are plotted along with the standard error.
	}
\end{figure*}

In order to answer \ref{rq_degree}, we investigated how many students successfully earn a degree in each major.
Figure~\ref{figure_degree_rates}a shows these degree-earning rates for all students in each major, while Fig.~\ref{figure_degree_rates}b shows these rates for female students and Fig.~\ref{figure_degree_rates}c for male students.
While these are broadly similar to an inverse of the drop rates in Fig.~\ref{figure_drop_plots}, since some students drop a major and subsequently declare the same major again, these degree-earning rates are a more direct measurement of persistence in a major.

Looking first at the overall rates in Fig.~\ref{figure_degree_rates}a, there are fairly wide differences across majors, from the lowest rate in physics of about 66\% to the highest in psychology and non-STEM, each at about 94\%.
The highest degree-earning rate in STEM occurs in computer science, with about 88\% of declared computer science majors completing the degree requirements.
As in Fig.~\ref{figure_drop_plots}, this can be at least partially explained by the requirements prior to declaring the major, which causes only students who have already progressed through a significant portion of the computer science curriculum to have declared a computer science major.

Considering then the differences for women (Fig.~\ref{figure_degree_rates}b) and men (Fig.~\ref{figure_degree_rates}c), we see relatively few gender differences in these degree-earning rates.
While the slightly higher completion rate of women in non-STEM and psychology or men in chemistry may turn out to be statistically significant, the effect sizes appear to be small due to differences of only 4-5\%.
As in Fig.~\ref{figure_drop_plots}, the largest difference between men and women seen here appears to be in physics, with 75\% of female physics majors earning a physics degree compared to 63\% of male physics majors.
However, the large error on these proportions, driven by the low sample size in physics shown in Fig.~\ref{figure_n_unique}, makes it difficult to draw any conclusions from this gender difference in physics degree-earning rates.
Similarly, women are more likely to complete a degree in economics, but again the size of the standard error prevents any conclusive statements about this difference.

Across all of Fig.~\ref{figure_degree_rates}, we note that since we have combined many majors for the ``non-STEM'' category, this is only a measure of the number of non-STEM majors who successfully earn a degree in any non-STEM major.
That is, a student who drops one non-STEM major but earns a degree in a different non-STEM major will still be counted as having successfully earned a non-STEM degree.
The same is true for the ``other STEM'' and ``engineering'' categories which also combine several majors.
The high ``success rates'' of computer science and psychology may be due in part to the structure of their program encouraging students to declare slightly later than other disciplines, and so this measure may not be capturing attrition that happens prior to an official declaration of major (e.g., a student intending to major in a discipline decides against it before ever declaring that major).
On the other hand, since all students enrolled in the engineering school are considered ``undeclared engineering,'' the relatively low degree-earning rate of engineering reflects attrition even from the first to the second term, which is not captured for many other majors in which most students have not yet formally declared a major in their first term.
Thus, each reported degree-earning rate here is a ceiling on the true rate that would include those students who intended to major but never declared.

\subsection{Mean GPA of Degree-Earners vs. Major-Droppers}
%RQ5: What are the GPA trends over time among students who earn a degree in a given major and those who drop that major? How do these trends differ for men and women?

\begin{figure*}
    \centering
    
	\includegraphics[width=0.95\textwidth]{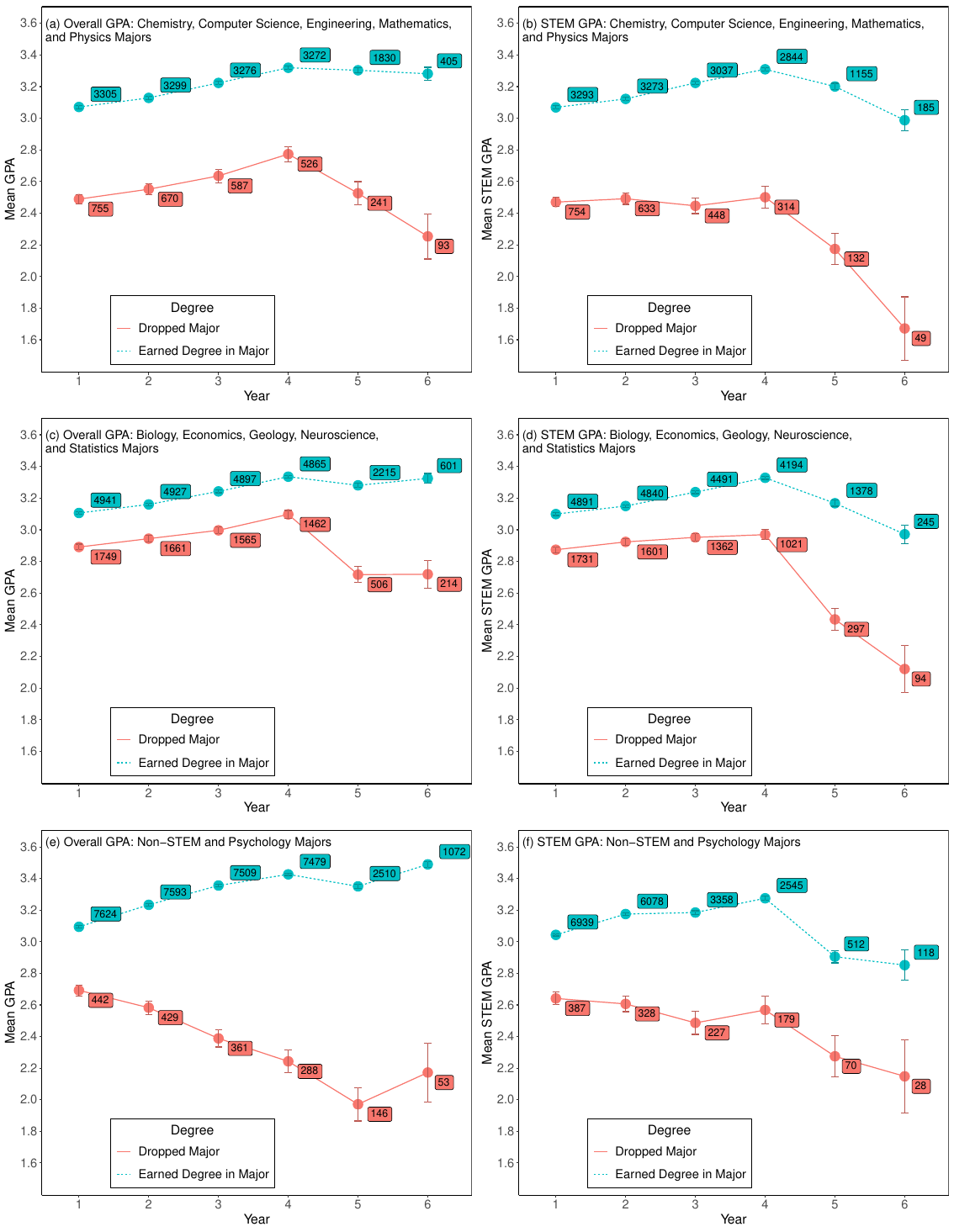}
	
	\caption{\label{figure_gpa_major}
	GPA and STEM GPA over time.
	Each GPA is calculated yearly, not cumulatively.
	Majors are divided into three groupings: (a) and (b) chemistry, computer science, engineering, mathematics, and physics; (c) and (d) biology, economics, geology, neuroscience, and statistics; and (e) and (f) non-STEM including psychology.
	GPA in all courses -- (a), (c), and (e) -- and in only STEM courses -- (b), (d), and (f) -- are calculated separately for two categories of students that declared at least one of the majors in each group: those that ultimately earned a degree in that group of majors and those that dropped from that group of majors.
	For each group, the mean GPA is plotted along with its standard error, with the sample size listed above each point and guides to the eye connecting the points.
	}
\end{figure*}

In order to further our understanding of why students may have dropped a given major and answer \ref{rq_gpa}, Fig.~\ref{figure_gpa_major} plots the mean GPA of students who declared different sets of majors and then either earned a degree within that set of majors or dropped those majors.
Note that students who dropped a major could have gone on to earn a degree with a different major or left the university without a degree.
Both overall GPA (Figs.~\ref{figure_gpa_major}a, \ref{figure_gpa_major}c, and \ref{figure_gpa_major}e) and STEM GPA (Figs.~\ref{figure_gpa_major}b, \ref{figure_gpa_major}d, and \ref{figure_gpa_major}f) are plotted.
Across all of Fig.~\ref{figure_gpa_major}, the large drop in sample size from year four to five and again from five to six is primarily due to students graduating.

We observe that in general, the students who drop that major have a lower GPA and STEM GPA than students who earned a degree in that major.
However, the difference between the two groups varies based on which cluster of majors we consider.
For chemistry, computer science, engineering, mathematics, and physics and astronomy majors (Figs.~\ref{figure_gpa_major}a and \ref{figure_gpa_major}b), those that earned a degree in this set of majors have a GPA of roughly 0.6 grade points higher than those that dropped.
This is consistent through the first four years of study, and similar between overall GPA and STEM GPA.
This difference in grade points represents a difference of one to two letter grades at the studied university, where, for example, the difference between a B and B$+$ is 0.25 grade points and between a B$+$ and A$-$ is 0.5 grade points.
Further, the number of students dropping from this set of majors is roughly 19\% of the total.

For biological science, economics, geology and environmental science, neuroscience, and statistics majors (Fig.~\ref{figure_gpa_major}c and \ref{figure_gpa_major}d), those that earned a degree in this set of majors have a GPA of roughly 0.2 grade points higher than those that dropped, with roughly 26\% of majors dropping.
As with the first set of majors, this is consistent between overall GPA and STEM GPA, and across the first four years of study.

Finally, for non-STEM majors including psychology (Fig.~\ref{figure_gpa_major}e and \ref{figure_gpa_major}f), the overall GPA disparity widens over time from roughly 0.4 grade points in the first year to roughly 1.2 grade points in the fourth year, while in STEM courses the GPA disparity rises from roughly 0.4 grade points in the first year to roughly 0.7 grade points in the fourth year.
Notably, a much smaller fraction of students are dropping from this set -- about 5\% of the total number of students -- which could be due in part to the wide net of considering all non-STEM majors.

\begin{figure*}
    \centering
    
	\includegraphics[width=0.95\textwidth]{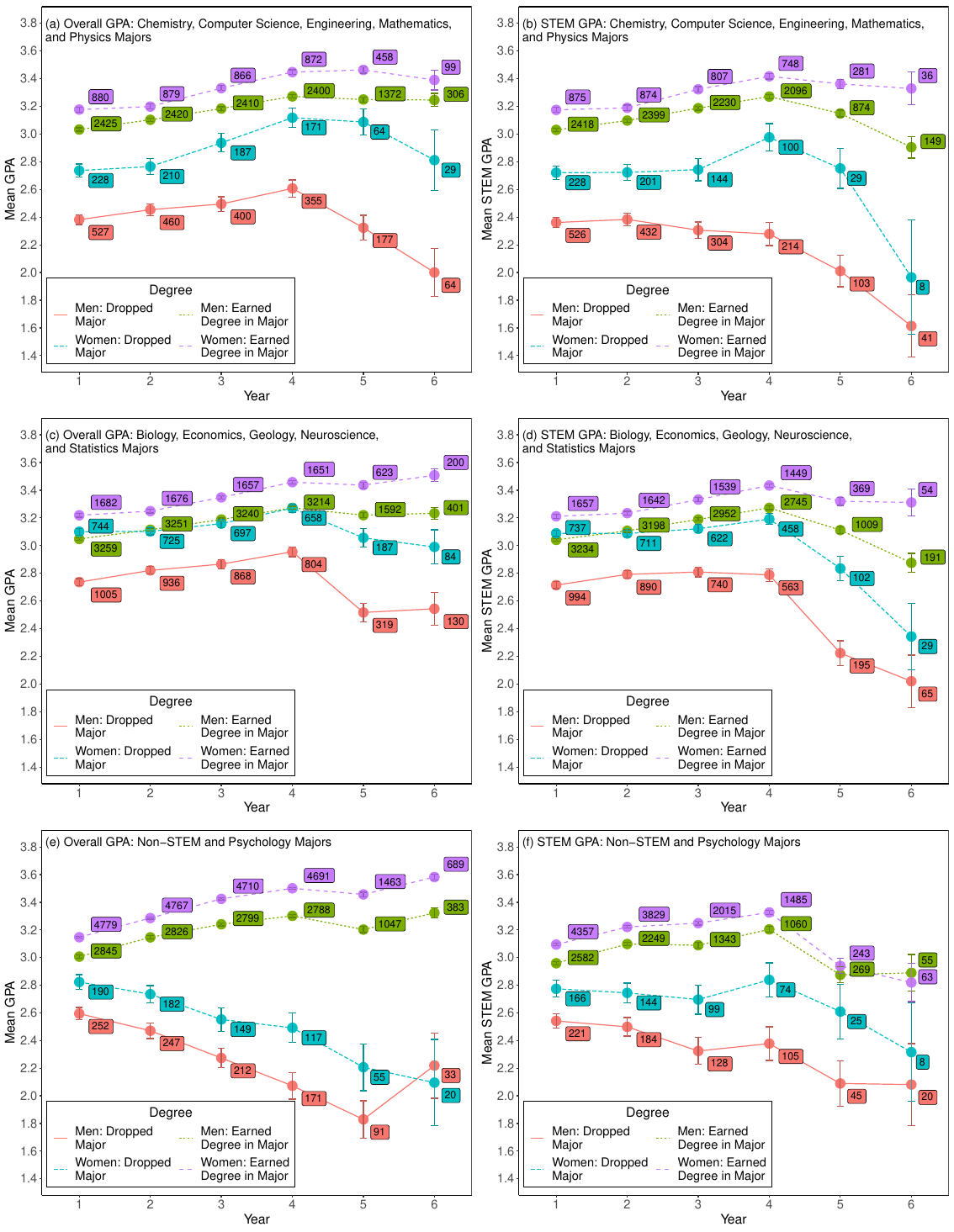}
	
	\caption{\label{figure_gender_gpa_major}
	GPA and STEM GPA over time by gender.
	Each GPA is calculated yearly, not cumulatively.
	Majors are divided into three groupings: (a) and (b) chemistry, computer science, engineering, mathematics, and physics; (c) and (d) biology, economics, geology, neuroscience, and statistics; and (e) and (f) non-STEM including psychology.
	GPA in all courses -- (a), (c), and (e) -- and in only STEM courses -- (b), (d), and (f) -- are calculated separately for four categories of students that declared at least one of the majors in each group: men and women that ultimately earned a degree in that group of majors and men and women that dropped from that group of majors.
	For each group, the mean GPA is plotted along with its standard error, with the sample size listed above each point and guides to the eye connecting the points.
	}
\end{figure*}

We further consider the same measures separately for men and women in Fig.~\ref{figure_gender_gpa_major}, with the same subfigure structure as Fig.~\ref{figure_gpa_major}.
That is, overall GPA is plotted in Figs.~\ref{figure_gender_gpa_major}a, \ref{figure_gender_gpa_major}c, and \ref{figure_gender_gpa_major}e and STEM GPA in Figs.~\ref{figure_gender_gpa_major}b, \ref{figure_gender_gpa_major}d, and \ref{figure_gender_gpa_major}f.
And students belonging to different clusters of majors are considered separately: chemistry, computer science, engineering, mathematics, and physics and astronomy majors are included in Figs.~\ref{figure_gender_gpa_major}a and \ref{figure_gender_gpa_major}b; biological science, economics, geology and environmental science, neuroscience, and statistics majors in Figs.~\ref{figure_gender_gpa_major}c and \ref{figure_gender_gpa_major}d; and non-STEM and psychology majors in Figs.~\ref{figure_gender_gpa_major}e and \ref{figure_gender_gpa_major}f.

In all cases, the main finding is that women are earning higher grades on average than men, both among those who drop a given major and those who earn a degree in that major.
However, the grade differences between men and women who dropped the major and men and women who earned a degree in the major are inconsistent.
In particular, we see that during the first four years in Fig.~\ref{figure_gender_gpa_major}b that among those who drop one of the listed STEM majors, women earn 0.4 to 0.6 grade points higher than men on average.
More specifically, these women have roughly a B$-$ average while the men have a C$+$ average.
Comparing this to those who earn a degree in those majors (Fig.~\ref{figure_gender_gpa_major}b), where women earn only 0.1 to 0.2 grade points higher than men on average, we see a trend where, on average, men will drop a major if they have significantly lower grades than their peers while women are dropping their major with only somewhat lower grades than their peers.

This trend is also reflected in the overall GPA of the same population (Fig.~\ref{figure_gender_gpa_major}a) as well as the overall and STEM GPA of the remainder of STEM majors (Figs.~\ref{figure_gender_gpa_major}c and \ref{figure_gender_gpa_major}d) and non-STEM majors including psychology (Figs.~\ref{figure_gender_gpa_major}e and \ref{figure_gender_gpa_major}f).
The trend is even more pronounced among students in the second cluster of STEM majors where in both overall GPA (Fig.~\ref{figure_gender_gpa_major}c) and STEM GPA (Fig.~\ref{figure_gender_gpa_major}d), the women who drop those majors have, on average, the same GPA as the men who earn a degree in those majors.

In order to test these differences, we report in Tables~\ref{table_gender_gpa_comparisons}a and \ref{table_gender_gpa_comparisons}b the results from four comparisons.
We compare the STEM GPA of the women who drop each set of STEM majors (Figs.~\ref{figure_gender_gpa_major}b and \ref{figure_gender_gpa_major}d) with the STEM GPA of men who both drop (Table~\ref{table_gender_gpa_comparisons}a) and earn a degree (Table~\ref{table_gender_gpa_comparisons}b) in the same majors.
These comparisons are performed with two measures: the $p$-value from a two-tailed $t$-test and the effect size from Cohen's $d$~\cite{freedman2007, cohen1988}.
These tests were not performed for non-STEM majors since such a low percentage of the students, roughly 5\%, dropped from the non-STEM majors (including psychology) altogether (Fig.~\ref{figure_gpa_major}).

Considering the STEM GPA of the first set of STEM majors in Table~\ref{table_gender_gpa_comparisons}a, namely chemistry, computer science, engineering, mathematics, and physics majors (corresponding to Fig.~\ref{figure_gender_gpa_major}b), we see again that the women who drop these majors are earning statistically significantly higher grades on average than the men who drop these majors ($p < 0.001$).
Further, Cohen's $d = 0.58$ indicates a medium effect size (i.e., $|d| \geq 0.50$)~\cite{cohen1988}, showing that the women who drop are earning meaningfully higher grades on average than the men who drop.
Table~\ref{table_gender_gpa_comparisons}b compares the women who drop with the men who persist and earn degrees in the majors, and we again see a statistically significant gender difference ($p < 0.001$) with a ``medium'' effect size ($d = -0.65$)~\cite{cohen1988}, now with the men having higher grades on average.
Notably, the effect sizes are similar in magnitude in both tables, which is consistent with these women who drop the major having an average STEM GPA roughly midway between the men who drop and the men who persist.

\begin{table*}
	\centering
	
	\begin{tabular}{c | C{0.90cm} C{0.90cm} C{0.90cm} | C{0.90cm} C{0.90cm} C{0.90cm} | C{1.2cm} C{0.90cm}}
		\multirow{2}{*}{\textbf{(a)}}	& \multicolumn{3}{c|}{Women who Dropped}	& \multicolumn{3}{c|}{Men who Dropped} & \multicolumn{2}{c}{Statistical} \\
										& \multicolumn{3}{c|}{the Major}			& \multicolumn{3}{c|}{the Major} & \multicolumn{2}{c}{Comparisons} \\
		Fig.	& $N$	& $M$	& $SD$			& $N$	& $M$	& $SD$	& $p$		& $d$			\\
		\hline
		\ref{figure_gender_gpa_major}b
				& 228	& 2.73	& 0.73			& 527	& 2.29	& 0.81 & $<0.001$	& 0.58			\\
		\ref{figure_gender_gpa_major}d
				& 744	& 3.10	& 0.64			& 1004	& 2.70	& 0.83 & $<0.001$	& 0.54			\\
	\end{tabular}
	
	\vspace{5mm}
	
	\begin{tabular}{c | C{0.90cm} C{0.90cm} C{0.90cm} | C{0.90cm} C{0.90cm} C{0.90cm} | C{1.2cm} C{0.90cm}}
		\multirow{2}{*}{\textbf{(b)}}	& \multicolumn{3}{c|}{Women who Dropped}	& \multicolumn{3}{c|}{Men who Earned} & \multicolumn{2}{c}{Statistical} \\
										& \multicolumn{3}{c|}{the Major}			& \multicolumn{3}{c|}{the Degree} & \multicolumn{2}{c}{Comparisons} \\
		Fig.	& $N$	& $M$	& $SD$			& $N$	& $M$	& $SD$	& $p$		& $d$ \\
		\hline
		\ref{figure_gender_gpa_major}b
				& 228	& 2.73	& 0.73			& 2425	& 3.14	& 0.52	& $<0.001$	& -0.65 \\
		\ref{figure_gender_gpa_major}d
				& 744	& 3.10	& 0.64			& 3257	& 3.15	& 0.52	& 0.047		& -0.09 \\
	\end{tabular}
	
	\caption{\label{table_gender_gpa_comparisons}
		A comparison of the cumulative STEM GPA of women who drop STEM majors with (a) the men who drop the same STEM majors and (b) the men who persist and earn a degree in the same STEM majors.
		Along with the number of students ($N$), we report the mean ($M$) and standard deviation ($SD$) of cumulative STEM GPA for each group.
		The $p$-value from a two-tailed $t$-test is reported comparing the women and men, along with Cohen's $d$ measuring the effect size of the gender difference (the sign of $d$ matches the sign of the mean GPA for women minus the mean GPA for men).
		The comparison is performed separately for two clusters of STEM majors corresponding to the indicated figure, i.e., for Fig.~\ref{figure_gender_gpa_major}b we consider chemistry, computer science, engineering, mathematics, and physics majors; for Fig.~\ref{figure_gender_gpa_major}d we consider biology, economics, geology, neuroscience, and statistics majors.
	}
\end{table*}

We turn then to the gendered STEM GPA differences of the other group of majors in Tables~\ref{table_gender_gpa_comparisons}a and \ref{table_gender_gpa_comparisons}b, namely biology, economics, geology, neuroscience, and statistics majors (corresponding to Fig.~\ref{figure_gender_gpa_major}d).
We see a similar pattern among the students who drop these majors (Table~\ref{table_gender_gpa_comparisons}a) as in the other set of majors: the women earn statistically significantly higher grades on average ($p < 0.001$) with a medium effect size ($d = 0.54$).
However, there is a notable difference when considering the men who earn degrees in these majors.
In particular, Table~\ref{table_gender_gpa_comparisons}b shows that the women who drop these majors are earning on average the same grades as men who persist, with $p = 0.047$ indicating only marginal statistical significance for an effect size ($d = -0.09$) that does not reach the threshold for ``small'' effect sizes (i.e., $|d| \geq 0.20$)~\cite{cohen1988}.
Thus, the women who are dropping these majors have, on average, the same grades as the men who persist and statistically significantly higher grades than the men who drop.

\section{Discussion}
%RQ1: \label{rq_declare} How many students major in each discipline? How many men and women major in each discipline?
%RQ2: \label{rq_drop} Do rates of attrition from the various majors differ? Do rates of attrition from the various majors differ for men and women?
%RQ3: \label{rq_droppers} Among those students who drop a given major, what degree, if any, do those students earn? How do these trends differ for men and women?
%RQ4: \label{rq_degree} What fraction of declared majors ultimately earn a degree in that major in each STEM subject area? How do these trends differ for men and women?
%RQ5: \label{rq_gpa} What are the GPA trends over time among students who earn a degree in a given major and those who drop that major? How do these trends differ for men and women?

In this section, we will begin by discussing the general trends (i.e., setting aside the gender differences), and then follow up with a discussion of the gender differences.

\subsection{General Enrollment Patterns}

Despite large differences in the number of students enrolling in different STEM disciplines at the studied university (Fig.~\ref{figure_n_unique}a), there are broadly similar patterns of when those students declare the major (Fig.~\ref{figure_peak_avg_term}a), with some some exceptions (i.e., engineering and computer science due to the constraints on when a student can declare a major).
However, there are notable differences in the attrition of students from the different majors (Fig.~\ref{figure_drop_plots}a), and the corresponding degree-earning rates (Fig.~\ref{figure_degree_rates}a).
Notably, a few STEM disciplines stand out as having particularly high rates of attrition (or low rates of degree completion for students who declared those majors),e.g., mathematics and physics.
This is consistent with the Leslie \textit{et al.} study which identifies mathematics and physics as the two STEM disciplines with the highest ``ability belief'' (i.e., emphasis on brilliance)~\cite{leslie2015}.
This trend of high attrition rates is particularly problematic for mathematics and physics, since these disciplines recruit very few students in the first place (Fig.~\ref{figure_n_unique}).
Moreover, mathematics and physics are also two disciplines with deeply hierarchical knowledge structures, which could influence student decision making, e.g., whether to leave the discipline after unsatisfactory experiences in earlier courses.

\subsection{Gendered Enrollment Patterns}

The most notable example of gender differences in enrollment patterns observed in our analysis is in Fig.~\ref{figure_percentages}.
In biological science, geological and environmental science, neuroscience, and statistics we see a balanced representation of men and women.
However, we see an underrepresentation of women in chemistry, computer science, engineering, mathematics, physics, and economics, and a corresponding overrepresentation of women in non-STEM including psychology (Fig.~\ref{figure_percentages}a).
Again, the results observed here are roughly consistent with those observed by Leslie \textit{et al.}~\cite{leslie2015}.
The gender imbalance in these STEM disciplines itself plays a pernicious role in recruitment and retention of women who do not have many role models and also affects the performance of women, who are constantly forced to prove themselves and counter the societal stereotypes working against them in these fields.

Furthermore, despite these differences in the number of men and women in these STEM disciplines, we see very few differences in the remainder of the enrollment measures, including the time of major-declaration (Figs.~\ref{figure_peak_avg_term}b and \ref{figure_peak_avg_term}c), rates of attrition (Figs.~\ref{figure_drop_plots}b and \ref{figure_drop_plots}c), and rates of degree-attainment (Figs.~\ref{figure_degree_rates}b and \ref{figure_degree_rates}c).
There are some hints towards differing rates of attrition in physics and economics but these differences suffer from large standard error, particularly in physics, due to a low sample size.
Moreover, we hypothesize that the higher attrition rate of men in physics and economics may be due to the pressure that women experience not to enroll in these disciplines in the first place, which would have occurred prior to the declaration of majors and instead manifests in the stark underrepresentation of women in these disciplines noted earlier (Fig.~\ref{figure_percentages}a).
There have been many studies that find highly problematic gender inequities in introductory physics and mathematics that could at least partly contribute to this underrepresentation~\cite{johnson2017, marshman2017, marshman2018, matz2017}.

\subsection{Trajectories of Students Who Drop a Major}

As with the attrition and degree-earning rates, we see broadly similar patterns between men and women who drop the various STEM majors (Fig.~\ref{figure_drop_plots_degree_by_group}).
One notable exceptions to this are that women who drop a mathematics major are more likely to instead earn a degree in economics than men who drop a mathematics major.
Moreover, we find that across all disciplines (including non-STEM majors) except psychology, women who drop a major are more likely to subsequently earn a degree in a different major at the same university than men who drop a major (see the ``No Degree'' entries in Fig.~\ref{figure_drop_plots_degree_by_group}), who are more likely to leave the university altogether, either by dropping out of college completely or transferring to another university.
This may be easier to see in Tables~\ref{table_appendix_M_supp} and \ref{table_appendix_F_supp} in Appendix C.

\subsection{Gendered GPA Differences}

We find pervasive and deeply troubling gendered trends in the overall GPA and STEM GPA of those students who drop different STEM majors.
In particular, Fig.~\ref{figure_gender_gpa_major} shows that the women who drop a STEM major have a higher average GPA than the men who drop a STEM major, and these differences are shown to be statistically significant as shown in Table~\ref{table_gender_gpa_comparisons}a.
This implies that on average, women are more likely to drop these majors with a significantly better GPA than men.
In chemistry, computer science, engineering, mathematics, and physics and astronomy (Fig.~\ref{figure_gender_gpa_major}a and \ref{figure_gender_gpa_major}b), the women dropping these majors have an average GPA (B$-$ to B) that is halfway between the men who drop these majors (C$+$ to B$-$) and the men who earn degrees with these majors (B to B$+$).
This trend is even worse in biology, economics, geological and environmental science, neuroscience, and statistics (Fig.~\ref{figure_gender_gpa_major}c and \ref{figure_gender_gpa_major}d), where the women who drop these majors have the same GPA as the men who earn degrees in these majors, both around B to B$+$ (see Table~\ref{table_gender_gpa_comparisons}b).
Thus, on average, among students with the same GPA, the women in all STEM majors are more likely to drop the majors than the men, who are more likely to persist.
It is important to note that this is even true among the STEM majors with gender imbalances in the population as well as those majors without imbalances.

While it is true that women at the studied university on average earn higher grades than men in most STEM courses, that does not explain why women are choosing to drop STEM majors with the same grades as men who persist.
The difference must then be coming from another source, for example inequitable and non-inclusive learning environments and lack of mentoring and support for women who may have lower self-efficacy~\cite{bandura1991, bandura1994, bandura1997, bandura1999, bandura2001, bandura2005}.
Women may also have a lower sense of belonging and value pertaining to remaining in these disciplines~\cite{eccles1984, eccles1990, eccles1994, safavian2019} if learning environments are not equitable and inclusive, especially because they must bear the high cost of managing the burden of societal stereotypes and the ensuing stereotype threats.
We hypothesize that the brilliance attributions of STEM disciplines~\cite{leslie2015} and who is likely to excel in them could influence women away from a discipline in which they could have succeeded.
Thus, without explicit effort to improve the learning environments in these disciplines, the culture and stereotypes surrounding STEM in general may be creating an environment in which women are being unfairly driven out of these fields in which they could have thrived while their male counterparts are not subjected to these same pressures and are persisting with worse performance.

\subsection{Limitations and Implications}
One limitation of this study is that physics which has the most consistently problematic gender differences also has the lowest number of students.
Future studies can make use of more data (either data available further back in time or as more data continues to accumulate) to explore the gender differences in physics better.
This study also limits its considerations to gender.
Other studies with larger data sets could investigate how other underserved populations are being left behind in STEM, such as underrepresented minority students, first-generation college students, or low-income students.

A critically important extension of this work would be for other institutions of different types and sizes to do similar analyses in order to broaden the wealth of knowledge available and continue to work towards the goal of equitable and inclusive education.
Other institutions noting similar highly problematic trends can help pinpoint common sources of inequities, while institutions that do not observe these trends may be able to identify how they have structured their programs to avoid these inequitable trends.
Studies such as this can thus provide a framework for other institutions to perform similar analyses, and for particular departments to understand how their own trends differ from those of other departments at their own university.
For instance, our findings here for, e.g., physics and mathematics drop out rates indicate that there is substantial room to improve their support of their intended majors so they do not drop out.
Focus on increasing equity and inclusion in learning is especially important in the early courses for these majors, since they are fraught with problematic gender differences and may be contributing to the underrepresentation of women in these majors in the first place.
Similar steps should be taken for the other majors that have low representation of women, especially computer science and engineering, and to a lesser extent chemistry.

The other STEM majors (biological sciences, economics, geology and environmental science, neuroscience, and statistics) should also take a closer look at which students are choosing to leave those disciplines, and why women choosing to leave have similar GPAs as the men who persist and earn degrees in those disciplines.
These are signs of inequity even in majors in which women are not underrepresented.
All of these issues should be addressed since they are critical for improving equity and inclusion in higher education STEM learning environments.

\section{Acknowledgments}
This research is supported by the National Science Foundation Grant DUE-1524575 and the Sloan Foundation Grant G-2018-11183.

\bibliography{refs}

\begin{thebibliography}{69}%
\makeatletter
\providecommand \@ifxundefined [1]{%
 \@ifx{#1\undefined}
}%
\providecommand \@ifnum [1]{%
 \ifnum #1\expandafter \@firstoftwo
 \else \expandafter \@secondoftwo
 \fi
}%
\providecommand \@ifx [1]{%
 \ifx #1\expandafter \@firstoftwo
 \else \expandafter \@secondoftwo
 \fi
}%
\providecommand \natexlab [1]{#1}%
\providecommand \enquote  [1]{``#1''}%
\providecommand \bibnamefont  [1]{#1}%
\providecommand \bibfnamefont [1]{#1}%
\providecommand \citenamefont [1]{#1}%
\providecommand \href@noop [0]{\@secondoftwo}%
\providecommand \href [0]{\begingroup \@sanitize@url \@href}%
\providecommand \@href[1]{\@@startlink{#1}\@@href}%
\providecommand \@@href[1]{\endgroup#1\@@endlink}%
\providecommand \@sanitize@url [0]{\catcode `\\12\catcode `\$12\catcode
  `\&12\catcode `\#12\catcode `\^12\catcode `\_12\catcode `\%12\relax}%
\providecommand \@@startlink[1]{}%
\providecommand \@@endlink[0]{}%
\providecommand \url  [0]{\begingroup\@sanitize@url \@url }%
\providecommand \@url [1]{\endgroup\@href {#1}{\urlprefix }}%
\providecommand \urlprefix  [0]{URL }%
\providecommand \Eprint [0]{\href }%
\providecommand \doibase [0]{https://doi.org/}%
\providecommand \selectlanguage [0]{\@gobble}%
\providecommand \bibinfo  [0]{\@secondoftwo}%
\providecommand \bibfield  [0]{\@secondoftwo}%
\providecommand \translation [1]{[#1]}%
\providecommand \BibitemOpen [0]{}%
\providecommand \bibitemStop [0]{}%
\providecommand \bibitemNoStop [0]{.\EOS\space}%
\providecommand \EOS [0]{\spacefactor3000\relax}%
\providecommand \BibitemShut  [1]{\csname bibitem#1\endcsname}%
\let\auto@bib@innerbib\@empty
%</preamble>
\bibitem [{\citenamefont {Johnson}(2012)}]{johnson2012}%
  \BibitemOpen
  \bibfield  {author} {\bibinfo {author} {\bibfnamefont {A.}~\bibnamefont
  {Johnson}},\ }\bibfield  {title} {\bibinfo {title} {The mathematics of sex:
  How biology and society conspire to limit talented women and girls},\
  }\href@noop {} {\bibfield  {journal} {\bibinfo  {journal} {Science
  Education}\ }\textbf {\bibinfo {volume} {96}},\ \bibinfo {pages} {960}
  (\bibinfo {year} {2012})}\BibitemShut {NoStop}%
\bibitem [{\citenamefont {Johnson}\ \emph {et~al.}(2017)\citenamefont
  {Johnson}, \citenamefont {Ong}, \citenamefont {Ko}, \citenamefont {Smith},\
  and\ \citenamefont {Hodari}}]{johnson2017}%
  \BibitemOpen
  \bibfield  {author} {\bibinfo {author} {\bibfnamefont {A.}~\bibnamefont
  {Johnson}}, \bibinfo {author} {\bibfnamefont {M.}~\bibnamefont {Ong}},
  \bibinfo {author} {\bibfnamefont {L.~T.}\ \bibnamefont {Ko}}, \bibinfo
  {author} {\bibfnamefont {J.}~\bibnamefont {Smith}},\ and\ \bibinfo {author}
  {\bibfnamefont {A.}~\bibnamefont {Hodari}},\ }\bibfield  {title} {\bibinfo
  {title} {Common challenges faced by women of color in physics, and actions
  faculty can take to minimize those challenges},\ }\href@noop {} {\bibfield
  {journal} {\bibinfo  {journal} {The Physics Teacher}\ }\textbf {\bibinfo
  {volume} {55}},\ \bibinfo {pages} {356} (\bibinfo {year} {2017})}\BibitemShut
  {NoStop}%
\bibitem [{\citenamefont {Metcalf}\ \emph {et~al.}(2018)\citenamefont
  {Metcalf}, \citenamefont {Russell},\ and\ \citenamefont
  {Hill}}]{metcalf2018}%
  \BibitemOpen
  \bibfield  {author} {\bibinfo {author} {\bibfnamefont {H.}~\bibnamefont
  {Metcalf}}, \bibinfo {author} {\bibfnamefont {D.}~\bibnamefont {Russell}},\
  and\ \bibinfo {author} {\bibfnamefont {C.}~\bibnamefont {Hill}},\ }\bibfield
  {title} {\bibinfo {title} {Broadening the science of broadening participation
  in stem through critical mixed methodologies and intersectionality
  frameworks},\ }\href {https://doi.org/10.1177/0002764218768872} {\bibfield
  {journal} {\bibinfo  {journal} {American Behavioral Scientist}\ }\textbf
  {\bibinfo {volume} {62}},\ \bibinfo {pages} {580} (\bibinfo {year} {2018})},\
  \Eprint {https://arxiv.org/abs/https://doi.org/10.1177/0002764218768872}
  {https://doi.org/10.1177/0002764218768872} \BibitemShut {NoStop}%
\bibitem [{\citenamefont {King}(2016)}]{king2016}%
  \BibitemOpen
  \bibfield  {author} {\bibinfo {author} {\bibfnamefont {B.}~\bibnamefont
  {King}},\ }\bibfield  {title} {\bibinfo {title} {Does postsecondary
  persistence in stem vary by gender?},\ }\href
  {https://doi.org/10.1177/2332858416669709} {\bibfield  {journal} {\bibinfo
  {journal} {AERA Open}\ }\textbf {\bibinfo {volume} {2}},\ \bibinfo {pages}
  {2332858416669709} (\bibinfo {year} {2016})},\ \Eprint
  {https://arxiv.org/abs/https://doi.org/10.1177/2332858416669709}
  {https://doi.org/10.1177/2332858416669709} \BibitemShut {NoStop}%
\bibitem [{\citenamefont {Maltese}\ and\ \citenamefont
  {Tai}(2011)}]{maltese2011}%
  \BibitemOpen
  \bibfield  {author} {\bibinfo {author} {\bibfnamefont {A.~V.}\ \bibnamefont
  {Maltese}}\ and\ \bibinfo {author} {\bibfnamefont {R.~H.}\ \bibnamefont
  {Tai}},\ }\bibfield  {title} {\bibinfo {title} {Pipeline persistence:
  Examining the association of educational experiences with earned degrees in
  {STEM} among {US} students},\ }\href@noop {} {\bibfield  {journal} {\bibinfo
  {journal} {Science Education}\ }\textbf {\bibinfo {volume} {95}},\ \bibinfo
  {pages} {877} (\bibinfo {year} {2011})}\BibitemShut {NoStop}%
\bibitem [{\citenamefont {Maltese}\ and\ \citenamefont
  {Cooper}(2017)}]{maltese2017}%
  \BibitemOpen
  \bibfield  {author} {\bibinfo {author} {\bibfnamefont {A.~V.}\ \bibnamefont
  {Maltese}}\ and\ \bibinfo {author} {\bibfnamefont {C.~S.}\ \bibnamefont
  {Cooper}},\ }\bibfield  {title} {\bibinfo {title} {Stem pathways: Do men and
  women differ in why they enter and exit?},\ }\href
  {https://doi.org/10.1177/2332858417727276} {\bibfield  {journal} {\bibinfo
  {journal} {AERA Open}\ }\textbf {\bibinfo {volume} {3}},\ \bibinfo {pages}
  {2332858417727276} (\bibinfo {year} {2017})},\ \Eprint
  {https://arxiv.org/abs/https://doi.org/10.1177/2332858417727276}
  {https://doi.org/10.1177/2332858417727276} \BibitemShut {NoStop}%
\bibitem [{\citenamefont {Means}\ \emph {et~al.}(2018)\citenamefont {Means},
  \citenamefont {Wang}, \citenamefont {Wei}, \citenamefont {Iwatani},\ and\
  \citenamefont {Peters}}]{means2018}%
  \BibitemOpen
  \bibfield  {author} {\bibinfo {author} {\bibfnamefont {B.}~\bibnamefont
  {Means}}, \bibinfo {author} {\bibfnamefont {H.}~\bibnamefont {Wang}},
  \bibinfo {author} {\bibfnamefont {X.}~\bibnamefont {Wei}}, \bibinfo {author}
  {\bibfnamefont {E.}~\bibnamefont {Iwatani}},\ and\ \bibinfo {author}
  {\bibfnamefont {V.}~\bibnamefont {Peters}},\ }\bibfield  {title} {\bibinfo
  {title} {Broadening participation in stem college majors: Effects of
  attending a stem-focused high school},\ }\href
  {https://doi.org/10.1177/2332858418806305} {\bibfield  {journal} {\bibinfo
  {journal} {AERA Open}\ }\textbf {\bibinfo {volume} {4}},\ \bibinfo {pages}
  {2332858418806305} (\bibinfo {year} {2018})},\ \Eprint
  {https://arxiv.org/abs/https://doi.org/10.1177/2332858418806305}
  {https://doi.org/10.1177/2332858418806305} \BibitemShut {NoStop}%
\bibitem [{\citenamefont {Borrego}\ \emph {et~al.}(2008)\citenamefont
  {Borrego}, \citenamefont {Streveler}, \citenamefont {Miller},\ and\
  \citenamefont {Smith}}]{borrego2008}%
  \BibitemOpen
  \bibfield  {author} {\bibinfo {author} {\bibfnamefont {M.}~\bibnamefont
  {Borrego}}, \bibinfo {author} {\bibfnamefont {R.~A.}\ \bibnamefont
  {Streveler}}, \bibinfo {author} {\bibfnamefont {R.~L.}\ \bibnamefont
  {Miller}},\ and\ \bibinfo {author} {\bibfnamefont {K.~A.}\ \bibnamefont
  {Smith}},\ }\bibfield  {title} {\bibinfo {title} {A new paradigm for a new
  field: {C}ommunicating representations of engineering education research},\
  }\href@noop {} {\bibfield  {journal} {\bibinfo  {journal} {Journal of
  Engineering Education}\ }\textbf {\bibinfo {volume} {97}},\ \bibinfo {pages}
  {147} (\bibinfo {year} {2008})}\BibitemShut {NoStop}%
\bibitem [{\citenamefont {Borrego}\ and\ \citenamefont
  {Bernhard}(2011)}]{borrego2011}%
  \BibitemOpen
  \bibfield  {author} {\bibinfo {author} {\bibfnamefont {M.}~\bibnamefont
  {Borrego}}\ and\ \bibinfo {author} {\bibfnamefont {J.}~\bibnamefont
  {Bernhard}},\ }\bibfield  {title} {\bibinfo {title} {The emergence of
  engineering education research as an internationally connected field of
  inquiry},\ }\href@noop {} {\bibfield  {journal} {\bibinfo  {journal} {Journal
  of Engineering Education}\ }\textbf {\bibinfo {volume} {100}},\ \bibinfo
  {pages} {14} (\bibinfo {year} {2011})}\BibitemShut {NoStop}%
\bibitem [{\citenamefont {Borrego}\ and\ \citenamefont
  {Henderson}(2014)}]{borrego2014}%
  \BibitemOpen
  \bibfield  {author} {\bibinfo {author} {\bibfnamefont {M.}~\bibnamefont
  {Borrego}}\ and\ \bibinfo {author} {\bibfnamefont {C.}~\bibnamefont
  {Henderson}},\ }\bibfield  {title} {\bibinfo {title} {Increasing the use of
  evidence-based teaching in {STEM} higher education: {A} comparison of eight
  change strategies},\ }\href@noop {} {\bibfield  {journal} {\bibinfo
  {journal} {Journal of Engineering Education}\ }\textbf {\bibinfo {volume}
  {103}},\ \bibinfo {pages} {220} (\bibinfo {year} {2014})}\BibitemShut
  {NoStop}%
\bibitem [{\citenamefont {Henderson}\ and\ \citenamefont
  {Dancy}(2008)}]{henderson2008}%
  \BibitemOpen
  \bibfield  {author} {\bibinfo {author} {\bibfnamefont {C.}~\bibnamefont
  {Henderson}}\ and\ \bibinfo {author} {\bibfnamefont {M.~H.}\ \bibnamefont
  {Dancy}},\ }\bibfield  {title} {\bibinfo {title} {Physics faculty and
  educational researchers: Divergent expectations as barriers to the diffusion
  of innovations},\ }\href@noop {} {\bibfield  {journal} {\bibinfo  {journal}
  {American Journal of Physics}\ }\textbf {\bibinfo {volume} {76}},\ \bibinfo
  {pages} {79} (\bibinfo {year} {2008})}\BibitemShut {NoStop}%
\bibitem [{\citenamefont {Dancy}\ and\ \citenamefont
  {Henderson}(2010)}]{dancy2010}%
  \BibitemOpen
  \bibfield  {author} {\bibinfo {author} {\bibfnamefont {M.}~\bibnamefont
  {Dancy}}\ and\ \bibinfo {author} {\bibfnamefont {C.}~\bibnamefont
  {Henderson}},\ }\bibfield  {title} {\bibinfo {title} {Pedagogical practices
  and instructional change of physics faculty},\ }\href@noop {} {\bibfield
  {journal} {\bibinfo  {journal} {American Journal of Physics}\ }\textbf
  {\bibinfo {volume} {78}},\ \bibinfo {pages} {1056} (\bibinfo {year}
  {2010})}\BibitemShut {NoStop}%
\bibitem [{\citenamefont {Henderson}\ \emph {et~al.}(2012)\citenamefont
  {Henderson}, \citenamefont {Dancy},\ and\ \citenamefont
  {Niewiadomska-Bugaj}}]{henderson2012}%
  \BibitemOpen
  \bibfield  {author} {\bibinfo {author} {\bibfnamefont {C.}~\bibnamefont
  {Henderson}}, \bibinfo {author} {\bibfnamefont {M.}~\bibnamefont {Dancy}},\
  and\ \bibinfo {author} {\bibfnamefont {M.}~\bibnamefont
  {Niewiadomska-Bugaj}},\ }\bibfield  {title} {\bibinfo {title} {Use of
  research-based instructional strategies in introductory physics: Where do
  faculty leave the innovation-decision process?},\ }\href
  {https://doi.org/10.1103/PhysRevSTPER.8.020104} {\bibfield  {journal}
  {\bibinfo  {journal} {Phys. Rev. ST Phys. Educ. Res.}\ }\textbf {\bibinfo
  {volume} {8}},\ \bibinfo {pages} {020104} (\bibinfo {year}
  {2012})}\BibitemShut {NoStop}%
\bibitem [{\citenamefont {{National Student Clearinghouse Research
  Center}}(2015)}]{nsc2015}%
  \BibitemOpen
  \bibfield  {author} {\bibinfo {author} {\bibnamefont {{National Student
  Clearinghouse Research Center}}},\ }\href
  {https://nscresearchcenter.org/snapshotreport-degreeattainment15/} {\emph
  {\bibinfo {title} {Science \& Engineering Degree Attainment: 2004 --
  2014}}},\ \bibinfo {type} {\unskip\space}\ (\bibinfo  {institution} {National
  Student Clearinghouse},\ \bibinfo {year} {2015})\ \bibinfo {note} {available
  at
  https://nscresearchcenter.org/snapshotreport-degreeattainment15/}\BibitemShut
  {NoStop}%
\bibitem [{\citenamefont {{National Science Board}}(2018)}]{nsf2018}%
  \BibitemOpen
  \bibfield  {author} {\bibinfo {author} {\bibnamefont {{National Science
  Board}}},\ }\href {https://www.nsf.gov/statistics/indicators/} {\emph
  {\bibinfo {title} {Science and Engineering Indicators 2018}}},\ \bibinfo
  {type} {\unskip\space}\ \bibinfo {number} {NSB-2018-1}\ (\bibinfo
  {institution} {National Science Foundation},\ \bibinfo {address} {Alexandria,
  VA},\ \bibinfo {year} {2018})\ \bibinfo {note} {available at
  https://www.nsf.gov/statistics/indicators/}\BibitemShut {NoStop}%
\bibitem [{\citenamefont {Baker}\ and\ \citenamefont
  {Inventado}(2014)}]{baker2014}%
  \BibitemOpen
  \bibfield  {author} {\bibinfo {author} {\bibfnamefont {R.~S.}\ \bibnamefont
  {Baker}}\ and\ \bibinfo {author} {\bibfnamefont {P.~S.}\ \bibnamefont
  {Inventado}},\ }\bibfield  {title} {\bibinfo {title} {Educational data mining
  and learning analytics},\ }in\ \href@noop {} {\emph {\bibinfo {booktitle}
  {Learning Analytics}}}\ (\bibinfo  {publisher} {Springer},\ \bibinfo {year}
  {2014})\ pp.\ \bibinfo {pages} {61--75}\BibitemShut {NoStop}%
\bibitem [{\citenamefont {Papamitsiou}\ and\ \citenamefont
  {Economides}(2014)}]{papamitsiou2014}%
  \BibitemOpen
  \bibfield  {author} {\bibinfo {author} {\bibfnamefont {Z.}~\bibnamefont
  {Papamitsiou}}\ and\ \bibinfo {author} {\bibfnamefont {A.~A.}\ \bibnamefont
  {Economides}},\ }\bibfield  {title} {\bibinfo {title} {Learning analytics and
  educational data mining in practice: A systematic literature review of
  empirical evidence},\ }\href@noop {} {\bibfield  {journal} {\bibinfo
  {journal} {Journal of Educational Technology \& Society}\ }\textbf {\bibinfo
  {volume} {17}},\ \bibinfo {pages} {49} (\bibinfo {year} {2014})}\BibitemShut
  {NoStop}%
\bibitem [{\citenamefont {Ohland}\ \emph {et~al.}(2008)\citenamefont {Ohland},
  \citenamefont {Sheppard}, \citenamefont {Lichtenstein}, \citenamefont {Eris},
  \citenamefont {Chachra},\ and\ \citenamefont {Layton}}]{ohland2008}%
  \BibitemOpen
  \bibfield  {author} {\bibinfo {author} {\bibfnamefont {M.~W.}\ \bibnamefont
  {Ohland}}, \bibinfo {author} {\bibfnamefont {S.~D.}\ \bibnamefont
  {Sheppard}}, \bibinfo {author} {\bibfnamefont {G.}~\bibnamefont
  {Lichtenstein}}, \bibinfo {author} {\bibfnamefont {O.}~\bibnamefont {Eris}},
  \bibinfo {author} {\bibfnamefont {D.}~\bibnamefont {Chachra}},\ and\ \bibinfo
  {author} {\bibfnamefont {R.~A.}\ \bibnamefont {Layton}},\ }\bibfield  {title}
  {\bibinfo {title} {Persistence, engagement, and migration in engineering
  programs},\ }\href@noop {} {\bibfield  {journal} {\bibinfo  {journal}
  {Journal of Engineering Education}\ }\textbf {\bibinfo {volume} {97}},\
  \bibinfo {pages} {259} (\bibinfo {year} {2008})}\BibitemShut {NoStop}%
\bibitem [{\citenamefont {Lord}\ \emph {et~al.}(2009)\citenamefont {Lord},
  \citenamefont {Camacho}, \citenamefont {Layton}, \citenamefont {Long},
  \citenamefont {Ohland},\ and\ \citenamefont {Wasburn}}]{lord2009}%
  \BibitemOpen
  \bibfield  {author} {\bibinfo {author} {\bibfnamefont {S.~M.}\ \bibnamefont
  {Lord}}, \bibinfo {author} {\bibfnamefont {M.~M.}\ \bibnamefont {Camacho}},
  \bibinfo {author} {\bibfnamefont {R.~A.}\ \bibnamefont {Layton}}, \bibinfo
  {author} {\bibfnamefont {R.~A.}\ \bibnamefont {Long}}, \bibinfo {author}
  {\bibfnamefont {M.~W.}\ \bibnamefont {Ohland}},\ and\ \bibinfo {author}
  {\bibfnamefont {M.~H.}\ \bibnamefont {Wasburn}},\ }\bibfield  {title}
  {\bibinfo {title} {Who's persisting in engineering? {A} comparative analysis
  of female and male {A}sian, black, {H}ispanic, {N}ative {A}merican, and white
  students},\ }\href@noop {} {\bibfield  {journal} {\bibinfo  {journal}
  {Journal of Women and Minorities in Science and Engineering}\ }\textbf
  {\bibinfo {volume} {15}},\ \bibinfo {pages} {167} (\bibinfo {year}
  {2009})}\BibitemShut {NoStop}%
\bibitem [{\citenamefont {Eris}\ \emph {et~al.}(2010)\citenamefont {Eris},
  \citenamefont {Chachra}, \citenamefont {Chen}, \citenamefont {Sheppard},
  \citenamefont {Ludlow}, \citenamefont {Rosca}, \citenamefont {Bailey},\ and\
  \citenamefont {Toye}}]{eris2010}%
  \BibitemOpen
  \bibfield  {author} {\bibinfo {author} {\bibfnamefont {O.}~\bibnamefont
  {Eris}}, \bibinfo {author} {\bibfnamefont {D.}~\bibnamefont {Chachra}},
  \bibinfo {author} {\bibfnamefont {H.~L.}\ \bibnamefont {Chen}}, \bibinfo
  {author} {\bibfnamefont {S.}~\bibnamefont {Sheppard}}, \bibinfo {author}
  {\bibfnamefont {L.}~\bibnamefont {Ludlow}}, \bibinfo {author} {\bibfnamefont
  {C.}~\bibnamefont {Rosca}}, \bibinfo {author} {\bibfnamefont
  {T.}~\bibnamefont {Bailey}},\ and\ \bibinfo {author} {\bibfnamefont
  {G.}~\bibnamefont {Toye}},\ }\bibfield  {title} {\bibinfo {title} {Outcomes
  of a longitudinal administration of the persistence in engineering survey},\
  }\href@noop {} {\bibfield  {journal} {\bibinfo  {journal} {Journal of
  Engineering Education}\ }\textbf {\bibinfo {volume} {99}},\ \bibinfo {pages}
  {371} (\bibinfo {year} {2010})}\BibitemShut {NoStop}%
\bibitem [{\citenamefont {Min}\ \emph {et~al.}(2011)\citenamefont {Min},
  \citenamefont {Zhang}, \citenamefont {Long}, \citenamefont {Anderson},\ and\
  \citenamefont {Ohland}}]{min2011}%
  \BibitemOpen
  \bibfield  {author} {\bibinfo {author} {\bibfnamefont {Y.}~\bibnamefont
  {Min}}, \bibinfo {author} {\bibfnamefont {G.}~\bibnamefont {Zhang}}, \bibinfo
  {author} {\bibfnamefont {R.~A.}\ \bibnamefont {Long}}, \bibinfo {author}
  {\bibfnamefont {T.~J.}\ \bibnamefont {Anderson}},\ and\ \bibinfo {author}
  {\bibfnamefont {M.~W.}\ \bibnamefont {Ohland}},\ }\bibfield  {title}
  {\bibinfo {title} {Nonparametric survival analysis of the loss rate of
  undergraduate engineering students},\ }\href@noop {} {\bibfield  {journal}
  {\bibinfo  {journal} {Journal of Engineering Education}\ }\textbf {\bibinfo
  {volume} {100}},\ \bibinfo {pages} {349} (\bibinfo {year}
  {2011})}\BibitemShut {NoStop}%
\bibitem [{\citenamefont {Lord}\ \emph {et~al.}(2015)\citenamefont {Lord},
  \citenamefont {Layton},\ and\ \citenamefont {Ohland}}]{lord2015}%
  \BibitemOpen
  \bibfield  {author} {\bibinfo {author} {\bibfnamefont {S.~M.}\ \bibnamefont
  {Lord}}, \bibinfo {author} {\bibfnamefont {R.~A.}\ \bibnamefont {Layton}},\
  and\ \bibinfo {author} {\bibfnamefont {M.~W.}\ \bibnamefont {Ohland}},\
  }\bibfield  {title} {\bibinfo {title} {Multi-institution study of student
  demographics and outcomes in electrical and computer engineering in the
  {USA}},\ }\href@noop {} {\bibfield  {journal} {\bibinfo  {journal} {IEEE
  Transactions on Education}\ }\textbf {\bibinfo {volume} {58}},\ \bibinfo
  {pages} {141} (\bibinfo {year} {2015})}\BibitemShut {NoStop}%
\bibitem [{\citenamefont {Ohland}\ and\ \citenamefont
  {Long}(2016)}]{ohland2016}%
  \BibitemOpen
  \bibfield  {author} {\bibinfo {author} {\bibfnamefont {M.~W.}\ \bibnamefont
  {Ohland}}\ and\ \bibinfo {author} {\bibfnamefont {R.~A.}\ \bibnamefont
  {Long}},\ }\bibfield  {title} {\bibinfo {title} {The multiple-institution
  database for investigating engineering longitudinal development: An
  experiential case study of data sharing and reuse},\ }\href@noop {}
  {\bibfield  {journal} {\bibinfo  {journal} {Advances in Engineering
  Education}\ }\textbf {\bibinfo {volume} {5}},\ \bibinfo {pages} {1} (\bibinfo
  {year} {2016})}\BibitemShut {NoStop}%
\bibitem [{\citenamefont {Matz}\ \emph {et~al.}(2017)\citenamefont {Matz},
  \citenamefont {Koester}, \citenamefont {Fiorini}, \citenamefont {Grom},
  \citenamefont {Shepard}, \citenamefont {Stangor}, \citenamefont {Weiner},\
  and\ \citenamefont {McKay}}]{matz2017}%
  \BibitemOpen
  \bibfield  {author} {\bibinfo {author} {\bibfnamefont {R.~L.}\ \bibnamefont
  {Matz}}, \bibinfo {author} {\bibfnamefont {B.~P.}\ \bibnamefont {Koester}},
  \bibinfo {author} {\bibfnamefont {S.}~\bibnamefont {Fiorini}}, \bibinfo
  {author} {\bibfnamefont {G.}~\bibnamefont {Grom}}, \bibinfo {author}
  {\bibfnamefont {L.}~\bibnamefont {Shepard}}, \bibinfo {author} {\bibfnamefont
  {C.~G.}\ \bibnamefont {Stangor}}, \bibinfo {author} {\bibfnamefont
  {B.}~\bibnamefont {Weiner}},\ and\ \bibinfo {author} {\bibfnamefont {T.~A.}\
  \bibnamefont {McKay}},\ }\bibfield  {title} {\bibinfo {title} {Patterns of
  gendered performance differences in large introductory courses at five
  research universities},\ }\href@noop {} {\bibfield  {journal} {\bibinfo
  {journal} {AERA Open}\ }\textbf {\bibinfo {volume} {3}},\ \bibinfo {pages}
  {1} (\bibinfo {year} {2017})}\BibitemShut {NoStop}%
\bibitem [{\citenamefont {Witherspoon}\ and\ \citenamefont
  {Schunn}(2019)}]{witherspoon2019}%
  \BibitemOpen
  \bibfield  {author} {\bibinfo {author} {\bibfnamefont {E.~B.}\ \bibnamefont
  {Witherspoon}}\ and\ \bibinfo {author} {\bibfnamefont {C.~D.}\ \bibnamefont
  {Schunn}},\ }\bibfield  {title} {\bibinfo {title} {Locating and understanding
  the largest gender differences in pathways to science degrees},\ }\href@noop
  {} {\bibfield  {journal} {\bibinfo  {journal} {Science Education}\ }\textbf
  {\bibinfo {volume} {104}},\ \bibinfo {pages} {144} (\bibinfo {year}
  {2019})}\BibitemShut {NoStop}%
\bibitem [{\citenamefont {Safavian}(2019)}]{safavian2019}%
  \BibitemOpen
  \bibfield  {author} {\bibinfo {author} {\bibfnamefont {N.}~\bibnamefont
  {Safavian}},\ }\bibfield  {title} {\bibinfo {title} {What makes them persist?
  expectancy-value beliefs and the math participation, performance, and
  preparedness of hispanic youth},\ }\href
  {https://doi.org/10.1177/2332858419869342} {\bibfield  {journal} {\bibinfo
  {journal} {AERA Open}\ }\textbf {\bibinfo {volume} {5}},\ \bibinfo {pages}
  {2332858419869342} (\bibinfo {year} {2019})},\ \Eprint
  {https://arxiv.org/abs/https://doi.org/10.1177/2332858419869342}
  {https://doi.org/10.1177/2332858419869342} \BibitemShut {NoStop}%
\bibitem [{\citenamefont {Gopalan}\ and\ \citenamefont
  {Nelson}(2019)}]{gopalan2019}%
  \BibitemOpen
  \bibfield  {author} {\bibinfo {author} {\bibfnamefont {M.}~\bibnamefont
  {Gopalan}}\ and\ \bibinfo {author} {\bibfnamefont {A.~A.}\ \bibnamefont
  {Nelson}},\ }\bibfield  {title} {\bibinfo {title} {Understanding the racial
  discipline gap in schools},\ }\href
  {https://doi.org/10.1177/2332858419844613} {\bibfield  {journal} {\bibinfo
  {journal} {AERA Open}\ }\textbf {\bibinfo {volume} {5}},\ \bibinfo {pages}
  {2332858419844613} (\bibinfo {year} {2019})},\ \Eprint
  {https://arxiv.org/abs/https://doi.org/10.1177/2332858419844613}
  {https://doi.org/10.1177/2332858419844613} \BibitemShut {NoStop}%
\bibitem [{\citenamefont {Crenshaw}\ \emph {et~al.}(1995)\citenamefont
  {Crenshaw}, \citenamefont {Gotanda}, \citenamefont {Peller},\ and\
  \citenamefont {Thomas}}]{crenshaw1995}%
  \BibitemOpen
  \bibfield  {author} {\bibinfo {author} {\bibfnamefont {K.}~\bibnamefont
  {Crenshaw}}, \bibinfo {author} {\bibfnamefont {N.}~\bibnamefont {Gotanda}},
  \bibinfo {author} {\bibfnamefont {G.}~\bibnamefont {Peller}},\ and\ \bibinfo
  {author} {\bibfnamefont {K.}~\bibnamefont {Thomas}},\ }\href@noop {} {\emph
  {\bibinfo {title} {Critical Race Theory: The Key Writings that Formed the
  Movement}}}\ (\bibinfo  {publisher} {The New Press},\ \bibinfo {year}
  {1995})\BibitemShut {NoStop}%
\bibitem [{\citenamefont {Kellner}(2003)}]{kellner2003}%
  \BibitemOpen
  \bibfield  {author} {\bibinfo {author} {\bibfnamefont {D.}~\bibnamefont
  {Kellner}},\ }\bibfield  {title} {\bibinfo {title} {Toward a critical theory
  of education},\ }\href@noop {} {\bibfield  {journal} {\bibinfo  {journal}
  {Democracy \& Nature}\ }\textbf {\bibinfo {volume} {9}},\ \bibinfo {pages}
  {51} (\bibinfo {year} {2003})}\BibitemShut {NoStop}%
\bibitem [{\citenamefont {Yosso}(2005)}]{yosso2005}%
  \BibitemOpen
  \bibfield  {author} {\bibinfo {author} {\bibfnamefont {T.~J.}\ \bibnamefont
  {Yosso}},\ }\bibfield  {title} {\bibinfo {title} {Whose culture has capital?
  {A} critical race theory discussion of community cultural wealth},\
  }\href@noop {} {\bibfield  {journal} {\bibinfo  {journal} {Race Ethnicity and
  Education}\ }\textbf {\bibinfo {volume} {8}},\ \bibinfo {pages} {69}
  (\bibinfo {year} {2005})}\BibitemShut {NoStop}%
\bibitem [{\citenamefont {Guti{\'e}rrez}(2009)}]{gutierrez2009}%
  \BibitemOpen
  \bibfield  {author} {\bibinfo {author} {\bibfnamefont {R.}~\bibnamefont
  {Guti{\'e}rrez}},\ }\bibfield  {title} {\bibinfo {title} {Framing equity:
  Helping students ``play the game'' and ``change the game''},\ }\href@noop {}
  {\bibfield  {journal} {\bibinfo  {journal} {Teaching for Excellence and
  Equity in Mathematics}\ }\textbf {\bibinfo {volume} {1}},\ \bibinfo {pages}
  {4} (\bibinfo {year} {2009})}\BibitemShut {NoStop}%
\bibitem [{\citenamefont {Taylor}\ \emph {et~al.}(2009)\citenamefont {Taylor},
  \citenamefont {Gillborn},\ and\ \citenamefont
  {Ladson-Billings}}]{taylor2009}%
  \BibitemOpen
  \bibfield  {author} {\bibinfo {author} {\bibfnamefont {E.}~\bibnamefont
  {Taylor}}, \bibinfo {author} {\bibfnamefont {D.}~\bibnamefont {Gillborn}},\
  and\ \bibinfo {author} {\bibfnamefont {G.}~\bibnamefont {Ladson-Billings}},\
  }\href@noop {} {\emph {\bibinfo {title} {Foundations of critical race theory
  in education}}}\ (\bibinfo  {publisher} {Routledge},\ \bibinfo {year}
  {2009})\BibitemShut {NoStop}%
\bibitem [{\citenamefont {Tolbert}\ \emph {et~al.}(2018)\citenamefont
  {Tolbert}, \citenamefont {Schindel},\ and\ \citenamefont
  {Rodriguez}}]{tolbert2018}%
  \BibitemOpen
  \bibfield  {author} {\bibinfo {author} {\bibfnamefont {S.}~\bibnamefont
  {Tolbert}}, \bibinfo {author} {\bibfnamefont {A.}~\bibnamefont {Schindel}},\
  and\ \bibinfo {author} {\bibfnamefont {A.~J.}\ \bibnamefont {Rodriguez}},\
  }\bibfield  {title} {\bibinfo {title} {Relevance and relational
  responsibility in justice-oriented science education research},\ }\href@noop
  {} {\bibfield  {journal} {\bibinfo  {journal} {Science Education}\ }\textbf
  {\bibinfo {volume} {102}},\ \bibinfo {pages} {796} (\bibinfo {year}
  {2018})}\BibitemShut {NoStop}%
\bibitem [{\citenamefont {Schenkel}\ and\ \citenamefont
  {Calabrese~Barton}(2020)}]{schenkel2020}%
  \BibitemOpen
  \bibfield  {author} {\bibinfo {author} {\bibfnamefont {K.}~\bibnamefont
  {Schenkel}}\ and\ \bibinfo {author} {\bibfnamefont {A.}~\bibnamefont
  {Calabrese~Barton}},\ }\bibfield  {title} {\bibinfo {title} {Critical science
  agency and power hierarchies: Restructuring power within groups to address
  injustice beyond them},\ }\href@noop {} {\bibfield  {journal} {\bibinfo
  {journal} {Science Education}\ } (\bibinfo {year} {2020})}\BibitemShut
  {NoStop}%
\bibitem [{\citenamefont {Bang}\ and\ \citenamefont {Medin}(2010)}]{bang2010}%
  \BibitemOpen
  \bibfield  {author} {\bibinfo {author} {\bibfnamefont {M.}~\bibnamefont
  {Bang}}\ and\ \bibinfo {author} {\bibfnamefont {D.}~\bibnamefont {Medin}},\
  }\bibfield  {title} {\bibinfo {title} {Cultural processes in science
  education: Supporting the navigation of multiple epistemologies},\
  }\href@noop {} {\bibfield  {journal} {\bibinfo  {journal} {Science
  Education}\ }\textbf {\bibinfo {volume} {94}},\ \bibinfo {pages} {1008}
  (\bibinfo {year} {2010})}\BibitemShut {NoStop}%
\bibitem [{\citenamefont {Estrada}\ \emph {et~al.}(2018)\citenamefont
  {Estrada}, \citenamefont {Eroy-Reveles},\ and\ \citenamefont
  {Matsui}}]{estrada2018}%
  \BibitemOpen
  \bibfield  {author} {\bibinfo {author} {\bibfnamefont {M.}~\bibnamefont
  {Estrada}}, \bibinfo {author} {\bibfnamefont {A.}~\bibnamefont
  {Eroy-Reveles}},\ and\ \bibinfo {author} {\bibfnamefont {J.}~\bibnamefont
  {Matsui}},\ }\bibfield  {title} {\bibinfo {title} {The influence of affirming
  kindness and community on broadening participation in {STEM} career
  pathways},\ }\href@noop {} {\bibfield  {journal} {\bibinfo  {journal} {Social
  Issues and Policy Review}\ }\textbf {\bibinfo {volume} {12}},\ \bibinfo
  {pages} {258} (\bibinfo {year} {2018})}\BibitemShut {NoStop}%
\bibitem [{\citenamefont {Ong}\ \emph {et~al.}(2018)\citenamefont {Ong},
  \citenamefont {Smith},\ and\ \citenamefont {Ko}}]{ong2018}%
  \BibitemOpen
  \bibfield  {author} {\bibinfo {author} {\bibfnamefont {M.}~\bibnamefont
  {Ong}}, \bibinfo {author} {\bibfnamefont {J.~M.}\ \bibnamefont {Smith}},\
  and\ \bibinfo {author} {\bibfnamefont {L.~T.}\ \bibnamefont {Ko}},\
  }\bibfield  {title} {\bibinfo {title} {Counterspaces for women of color in
  {STEM} higher education: Marginal and central spaces for persistence and
  success},\ }\href@noop {} {\bibfield  {journal} {\bibinfo  {journal} {Journal
  of Research in Science Teaching}\ }\textbf {\bibinfo {volume} {55}},\
  \bibinfo {pages} {206} (\bibinfo {year} {2018})}\BibitemShut {NoStop}%
\bibitem [{\citenamefont {Seron}\ \emph {et~al.}(2016)\citenamefont {Seron},
  \citenamefont {Silbey}, \citenamefont {Cech},\ and\ \citenamefont
  {Rubineau}}]{seron2016}%
  \BibitemOpen
  \bibfield  {author} {\bibinfo {author} {\bibfnamefont {C.}~\bibnamefont
  {Seron}}, \bibinfo {author} {\bibfnamefont {S.~S.}\ \bibnamefont {Silbey}},
  \bibinfo {author} {\bibfnamefont {E.}~\bibnamefont {Cech}},\ and\ \bibinfo
  {author} {\bibfnamefont {B.}~\bibnamefont {Rubineau}},\ }\bibfield  {title}
  {\bibinfo {title} {Persistence is cultural: Professional socialization and
  the reproduction of sex segregation},\ }\href@noop {} {\bibfield  {journal}
  {\bibinfo  {journal} {Work and Occupations}\ }\textbf {\bibinfo {volume}
  {43}},\ \bibinfo {pages} {178} (\bibinfo {year} {2016})}\BibitemShut
  {NoStop}%
\bibitem [{\citenamefont {Brawner}\ \emph {et~al.}(2012)\citenamefont
  {Brawner}, \citenamefont {Camacho}, \citenamefont {Lord}, \citenamefont
  {Long},\ and\ \citenamefont {Ohland}}]{brawner2012}%
  \BibitemOpen
  \bibfield  {author} {\bibinfo {author} {\bibfnamefont {C.~E.}\ \bibnamefont
  {Brawner}}, \bibinfo {author} {\bibfnamefont {M.~M.}\ \bibnamefont
  {Camacho}}, \bibinfo {author} {\bibfnamefont {S.~M.}\ \bibnamefont {Lord}},
  \bibinfo {author} {\bibfnamefont {R.~A.}\ \bibnamefont {Long}},\ and\
  \bibinfo {author} {\bibfnamefont {M.~W.}\ \bibnamefont {Ohland}},\ }\bibfield
   {title} {\bibinfo {title} {Women in industrial engineering: Stereotypes,
  persistence, and perspectives},\ }\href@noop {} {\bibfield  {journal}
  {\bibinfo  {journal} {Journal of Engineering Education}\ }\textbf {\bibinfo
  {volume} {101}},\ \bibinfo {pages} {288} (\bibinfo {year}
  {2012})}\BibitemShut {NoStop}%
\bibitem [{\citenamefont {Brawner}\ \emph {et~al.}(2015)\citenamefont
  {Brawner}, \citenamefont {Lord}, \citenamefont {Layton}, \citenamefont
  {Ohland},\ and\ \citenamefont {Long}}]{brawner2015}%
  \BibitemOpen
  \bibfield  {author} {\bibinfo {author} {\bibfnamefont {C.~E.}\ \bibnamefont
  {Brawner}}, \bibinfo {author} {\bibfnamefont {S.~M.}\ \bibnamefont {Lord}},
  \bibinfo {author} {\bibfnamefont {R.~A.}\ \bibnamefont {Layton}}, \bibinfo
  {author} {\bibfnamefont {M.~W.}\ \bibnamefont {Ohland}},\ and\ \bibinfo
  {author} {\bibfnamefont {R.~A.}\ \bibnamefont {Long}},\ }\bibfield  {title}
  {\bibinfo {title} {Factors affecting women's persistence in chemical
  engineering},\ }\href@noop {} {\bibfield  {journal} {\bibinfo  {journal}
  {International Journal of Engineering Education}\ }\textbf {\bibinfo {volume}
  {31}},\ \bibinfo {pages} {1431} (\bibinfo {year} {2015})}\BibitemShut
  {NoStop}%
\bibitem [{\citenamefont {Ganley}\ \emph {et~al.}(2018)\citenamefont {Ganley},
  \citenamefont {George}, \citenamefont {Cimpian},\ and\ \citenamefont
  {Makowski}}]{ganley2018}%
  \BibitemOpen
  \bibfield  {author} {\bibinfo {author} {\bibfnamefont {C.~M.}\ \bibnamefont
  {Ganley}}, \bibinfo {author} {\bibfnamefont {C.~E.}\ \bibnamefont {George}},
  \bibinfo {author} {\bibfnamefont {J.~R.}\ \bibnamefont {Cimpian}},\ and\
  \bibinfo {author} {\bibfnamefont {M.~B.}\ \bibnamefont {Makowski}},\
  }\bibfield  {title} {\bibinfo {title} {Gender equity in college majors:
  Looking beyond the stem/non-stem dichotomy for answers regarding female
  participation},\ }\href {https://doi.org/10.3102/0002831217740221} {\bibfield
   {journal} {\bibinfo  {journal} {American Educational Research Journal}\
  }\textbf {\bibinfo {volume} {55}},\ \bibinfo {pages} {453} (\bibinfo {year}
  {2018})},\ \Eprint
  {https://arxiv.org/abs/https://doi.org/10.3102/0002831217740221}
  {https://doi.org/10.3102/0002831217740221} \BibitemShut {NoStop}%
\bibitem [{\citenamefont {Leslie}\ \emph {et~al.}(2015)\citenamefont {Leslie},
  \citenamefont {Cimpian}, \citenamefont {Meyer},\ and\ \citenamefont
  {Freeland}}]{leslie2015}%
  \BibitemOpen
  \bibfield  {author} {\bibinfo {author} {\bibfnamefont {S.-J.}\ \bibnamefont
  {Leslie}}, \bibinfo {author} {\bibfnamefont {A.}~\bibnamefont {Cimpian}},
  \bibinfo {author} {\bibfnamefont {M.}~\bibnamefont {Meyer}},\ and\ \bibinfo
  {author} {\bibfnamefont {E.}~\bibnamefont {Freeland}},\ }\bibfield  {title}
  {\bibinfo {title} {Expectations of brilliance underlie gender distributions
  across academic disciplines},\ }\href@noop {} {\bibfield  {journal} {\bibinfo
   {journal} {Science}\ }\textbf {\bibinfo {volume} {347}},\ \bibinfo {pages}
  {262} (\bibinfo {year} {2015})}\BibitemShut {NoStop}%
\bibitem [{\citenamefont {Bian}\ \emph {et~al.}(2017)\citenamefont {Bian},
  \citenamefont {Leslie},\ and\ \citenamefont {Cimpian}}]{bian2017}%
  \BibitemOpen
  \bibfield  {author} {\bibinfo {author} {\bibfnamefont {L.}~\bibnamefont
  {Bian}}, \bibinfo {author} {\bibfnamefont {S.-J.}\ \bibnamefont {Leslie}},\
  and\ \bibinfo {author} {\bibfnamefont {A.}~\bibnamefont {Cimpian}},\
  }\bibfield  {title} {\bibinfo {title} {Gender stereotypes about intellectual
  ability emerge early and influence children's interests},\ }\href@noop {}
  {\bibfield  {journal} {\bibinfo  {journal} {Science}\ }\textbf {\bibinfo
  {volume} {355}},\ \bibinfo {pages} {389} (\bibinfo {year}
  {2017})}\BibitemShut {NoStop}%
\bibitem [{\citenamefont {Bian}(2017)}]{bian2017thesis}%
  \BibitemOpen
  \bibfield  {author} {\bibinfo {author} {\bibfnamefont {L.}~\bibnamefont
  {Bian}},\ }\emph {\bibinfo {title} {The roots of gender gaps: investigating
  the development of gender stereotypes about intelligence}},\ \href@noop {}
  {Ph.D. thesis},\ \bibinfo  {school} {University of Illinois at
  Urbana-Champaign} (\bibinfo {year} {2017})\BibitemShut {NoStop}%
\bibitem [{\citenamefont {Bian}\ \emph
  {et~al.}(2018{\natexlab{a}})\citenamefont {Bian}, \citenamefont {Leslie},
  \citenamefont {Murphy},\ and\ \citenamefont {Cimpian}}]{bian2018messages}%
  \BibitemOpen
  \bibfield  {author} {\bibinfo {author} {\bibfnamefont {L.}~\bibnamefont
  {Bian}}, \bibinfo {author} {\bibfnamefont {S.-J.}\ \bibnamefont {Leslie}},
  \bibinfo {author} {\bibfnamefont {M.~C.}\ \bibnamefont {Murphy}},\ and\
  \bibinfo {author} {\bibfnamefont {A.}~\bibnamefont {Cimpian}},\ }\bibfield
  {title} {\bibinfo {title} {Messages about brilliance undermine women's
  interest in educational and professional opportunities},\ }\href@noop {}
  {\bibfield  {journal} {\bibinfo  {journal} {Journal of Experimental Social
  Psychology}\ }\textbf {\bibinfo {volume} {76}},\ \bibinfo {pages} {404}
  (\bibinfo {year} {2018}{\natexlab{a}})}\BibitemShut {NoStop}%
\bibitem [{\citenamefont {Bian}\ \emph
  {et~al.}(2018{\natexlab{b}})\citenamefont {Bian}, \citenamefont {Leslie},\
  and\ \citenamefont {Cimpian}}]{bian2018evidence}%
  \BibitemOpen
  \bibfield  {author} {\bibinfo {author} {\bibfnamefont {L.}~\bibnamefont
  {Bian}}, \bibinfo {author} {\bibfnamefont {S.-J.}\ \bibnamefont {Leslie}},\
  and\ \bibinfo {author} {\bibfnamefont {A.}~\bibnamefont {Cimpian}},\
  }\bibfield  {title} {\bibinfo {title} {Evidence of bias against girls and
  women in contexts that emphasize intellectual ability},\ }\href@noop {}
  {\bibfield  {journal} {\bibinfo  {journal} {American Psychologist}\ }\textbf
  {\bibinfo {volume} {73}},\ \bibinfo {pages} {1139} (\bibinfo {year}
  {2018}{\natexlab{b}})}\BibitemShut {NoStop}%
\bibitem [{\citenamefont {Eccles~(Parsons)}\ \emph {et~al.}(1984)\citenamefont
  {Eccles~(Parsons)}, \citenamefont {Adler},\ and\ \citenamefont
  {Meece}}]{eccles1984}%
  \BibitemOpen
  \bibfield  {author} {\bibinfo {author} {\bibfnamefont {J.}~\bibnamefont
  {Eccles~(Parsons)}}, \bibinfo {author} {\bibfnamefont {T.}~\bibnamefont
  {Adler}},\ and\ \bibinfo {author} {\bibfnamefont {J.}~\bibnamefont {Meece}},\
  }\bibfield  {title} {\bibinfo {title} {Sex differences in achievement: A test
  of alternative theories},\ }\href@noop {} {\bibfield  {journal} {\bibinfo
  {journal} {Journal of Personality and Social Psychology}\ }\textbf {\bibinfo
  {volume} {46}},\ \bibinfo {pages} {26} (\bibinfo {year} {1984})}\BibitemShut
  {NoStop}%
\bibitem [{\citenamefont {Eccles}\ \emph {et~al.}(1990)\citenamefont {Eccles},
  \citenamefont {Jacobs},\ and\ \citenamefont {Harold}}]{eccles1990}%
  \BibitemOpen
  \bibfield  {author} {\bibinfo {author} {\bibfnamefont {J.~S.}\ \bibnamefont
  {Eccles}}, \bibinfo {author} {\bibfnamefont {J.~E.}\ \bibnamefont {Jacobs}},\
  and\ \bibinfo {author} {\bibfnamefont {R.~D.}\ \bibnamefont {Harold}},\
  }\bibfield  {title} {\bibinfo {title} {Gender role stereotypes, expectancy
  effects, and parents' socialization of gender differences},\ }\href@noop {}
  {\bibfield  {journal} {\bibinfo  {journal} {Journal of Social Issues}\
  }\textbf {\bibinfo {volume} {46}},\ \bibinfo {pages} {183} (\bibinfo {year}
  {1990})}\BibitemShut {NoStop}%
\bibitem [{\citenamefont {Eccles}(1994)}]{eccles1994}%
  \BibitemOpen
  \bibfield  {author} {\bibinfo {author} {\bibfnamefont {J.~S.}\ \bibnamefont
  {Eccles}},\ }\bibfield  {title} {\bibinfo {title} {Understanding women's
  educational and occupational choices: Applying the {E}ccles et al. model of
  achievement-related choices},\ }\href@noop {} {\bibfield  {journal} {\bibinfo
   {journal} {Psychology of Women Quarterly}\ }\textbf {\bibinfo {volume}
  {18}},\ \bibinfo {pages} {585} (\bibinfo {year} {1994})}\BibitemShut
  {NoStop}%
\bibitem [{\citenamefont {Bandura}(1991)}]{bandura1991}%
  \BibitemOpen
  \bibfield  {author} {\bibinfo {author} {\bibfnamefont {A.}~\bibnamefont
  {Bandura}},\ }\bibfield  {title} {\bibinfo {title} {Social cognitive theory
  of self-regulation},\ }\href@noop {} {\bibfield  {journal} {\bibinfo
  {journal} {Organizational Behavior and Human Decision Processes}\ }\textbf
  {\bibinfo {volume} {50}},\ \bibinfo {pages} {248} (\bibinfo {year}
  {1991})}\BibitemShut {NoStop}%
\bibitem [{\citenamefont {Bandura}(1994)}]{bandura1994}%
  \BibitemOpen
  \bibfield  {author} {\bibinfo {author} {\bibfnamefont {A.}~\bibnamefont
  {Bandura}},\ }\bibfield  {title} {\bibinfo {title} {Self-efficacy},\ }in\
  \href@noop {} {\emph {\bibinfo {booktitle} {Encyclopedia of Psychology}}},\
  Vol.~\bibinfo {volume} {3},\ \bibinfo {editor} {edited by\ \bibinfo {editor}
  {\bibfnamefont {R.~J.}\ \bibnamefont {Corsini}}}\ (\bibinfo  {publisher}
  {Wiley},\ \bibinfo {year} {1994})\ pp.\ \bibinfo {pages}
  {368--369}\BibitemShut {NoStop}%
\bibitem [{\citenamefont {Bandura}(1997)}]{bandura1997}%
  \BibitemOpen
  \bibfield  {author} {\bibinfo {author} {\bibfnamefont {A.}~\bibnamefont
  {Bandura}},\ }\href@noop {} {\emph {\bibinfo {title} {Self-efficacy: {T}he
  Exercise of Control}}}\ (\bibinfo  {publisher} {Macmillan},\ \bibinfo {year}
  {1997})\BibitemShut {NoStop}%
\bibitem [{\citenamefont {Bandura}(1999)}]{bandura1999}%
  \BibitemOpen
  \bibfield  {author} {\bibinfo {author} {\bibfnamefont {A.}~\bibnamefont
  {Bandura}},\ }\bibfield  {title} {\bibinfo {title} {Social cognitive theory
  of personality},\ }in\ \href@noop {} {\emph {\bibinfo {booktitle} {Handbook
  of Personality}}},\ Vol.~\bibinfo {volume} {2},\ \bibinfo {editor} {edited
  by\ \bibinfo {editor} {\bibfnamefont {L.~A.}\ \bibnamefont {Pervin}}\ and\
  \bibinfo {editor} {\bibfnamefont {O.~P.}\ \bibnamefont {John}}}\ (\bibinfo
  {publisher} {Oxford University Press},\ \bibinfo {year} {1999})\ pp.\
  \bibinfo {pages} {154--196}\BibitemShut {NoStop}%
\bibitem [{\citenamefont {Bandura}(2001)}]{bandura2001}%
  \BibitemOpen
  \bibfield  {author} {\bibinfo {author} {\bibfnamefont {A.}~\bibnamefont
  {Bandura}},\ }\bibfield  {title} {\bibinfo {title} {Social cognitive theory:
  An agentic perspective},\ }\href@noop {} {\bibfield  {journal} {\bibinfo
  {journal} {Annual Review of Psychology}\ }\textbf {\bibinfo {volume} {52}},\
  \bibinfo {pages} {1} (\bibinfo {year} {2001})}\BibitemShut {NoStop}%
\bibitem [{\citenamefont {Bandura}(2005)}]{bandura2005}%
  \BibitemOpen
  \bibfield  {author} {\bibinfo {author} {\bibfnamefont {A.}~\bibnamefont
  {Bandura}},\ }\bibfield  {title} {\bibinfo {title} {The evolution of social
  cognitive theory},\ }in\ \href@noop {} {\emph {\bibinfo {booktitle} {Great
  Minds in Management}}},\ \bibinfo {editor} {edited by\ \bibinfo {editor}
  {\bibfnamefont {K.~G.}\ \bibnamefont {Smith}}\ and\ \bibinfo {editor}
  {\bibfnamefont {M.~A.}\ \bibnamefont {Hitt}}}\ (\bibinfo  {publisher} {Oxford
  University Press},\ \bibinfo {year} {2005})\ pp.\ \bibinfo {pages}
  {9--35}\BibitemShut {NoStop}%
\bibitem [{\citenamefont {Astin}(1993)}]{astin1993}%
  \BibitemOpen
  \bibfield  {author} {\bibinfo {author} {\bibfnamefont {A.~W.}\ \bibnamefont
  {Astin}},\ }\href@noop {} {\emph {\bibinfo {title} {What Matters in
  College}}},\ Vol.~\bibinfo {volume} {9}\ (\bibinfo  {publisher}
  {Jossey-Bass},\ \bibinfo {year} {1993})\BibitemShut {NoStop}%
\bibitem [{\citenamefont {Felder}\ \emph {et~al.}(1995)\citenamefont {Felder},
  \citenamefont {Felder}, \citenamefont {Mauney}, \citenamefont {Hamrin~Jr.},\
  and\ \citenamefont {Dietz}}]{felder1995}%
  \BibitemOpen
  \bibfield  {author} {\bibinfo {author} {\bibfnamefont {R.~M.}\ \bibnamefont
  {Felder}}, \bibinfo {author} {\bibfnamefont {G.~N.}\ \bibnamefont {Felder}},
  \bibinfo {author} {\bibfnamefont {M.}~\bibnamefont {Mauney}}, \bibinfo
  {author} {\bibfnamefont {C.~E.}\ \bibnamefont {Hamrin~Jr.}},\ and\ \bibinfo
  {author} {\bibfnamefont {E.~J.}\ \bibnamefont {Dietz}},\ }\bibfield  {title}
  {\bibinfo {title} {A longitudinal study of engineering student performance
  and retention. {III}. {Gender} differences in student performance and
  attitudes},\ }\href@noop {} {\bibfield  {journal} {\bibinfo  {journal}
  {Journal of Engineering Education}\ }\textbf {\bibinfo {volume} {84}},\
  \bibinfo {pages} {151} (\bibinfo {year} {1995})}\BibitemShut {NoStop}%
\bibitem [{\citenamefont {Felder}\ \emph {et~al.}(1998)\citenamefont {Felder},
  \citenamefont {Felder},\ and\ \citenamefont {Dietz}}]{felder1998}%
  \BibitemOpen
  \bibfield  {author} {\bibinfo {author} {\bibfnamefont {R.~M.}\ \bibnamefont
  {Felder}}, \bibinfo {author} {\bibfnamefont {G.~N.}\ \bibnamefont {Felder}},\
  and\ \bibinfo {author} {\bibfnamefont {E.~J.}\ \bibnamefont {Dietz}},\
  }\bibfield  {title} {\bibinfo {title} {A longitudinal study of engineering
  student performance and retention. {V}. {Comparisons} with
  traditionally-taught students},\ }\href@noop {} {\bibfield  {journal}
  {\bibinfo  {journal} {Journal of Engineering Education}\ }\textbf {\bibinfo
  {volume} {87}},\ \bibinfo {pages} {469} (\bibinfo {year} {1998})}\BibitemShut
  {NoStop}%
\bibitem [{\citenamefont {Cheryan}\ \emph {et~al.}(2017)\citenamefont
  {Cheryan}, \citenamefont {Ziegler}, \citenamefont {Montoya},\ and\
  \citenamefont {Jiang}}]{cheryan2017}%
  \BibitemOpen
  \bibfield  {author} {\bibinfo {author} {\bibfnamefont {S.}~\bibnamefont
  {Cheryan}}, \bibinfo {author} {\bibfnamefont {S.~A.}\ \bibnamefont
  {Ziegler}}, \bibinfo {author} {\bibfnamefont {A.~K.}\ \bibnamefont
  {Montoya}},\ and\ \bibinfo {author} {\bibfnamefont {L.}~\bibnamefont
  {Jiang}},\ }\bibfield  {title} {\bibinfo {title} {Why are some {STEM} fields
  more gender balanced than others?},\ }\href@noop {} {\bibfield  {journal}
  {\bibinfo  {journal} {Psychological Bulletin}\ }\textbf {\bibinfo {volume}
  {143}},\ \bibinfo {pages} {1} (\bibinfo {year} {2017})}\BibitemShut {NoStop}%
\bibitem [{\citenamefont {Bianchini}\ \emph {et~al.}(2002)\citenamefont
  {Bianchini}, \citenamefont {Whitney}, \citenamefont {Breton},\ and\
  \citenamefont {Hilton-Brown}}]{bianchini2002}%
  \BibitemOpen
  \bibfield  {author} {\bibinfo {author} {\bibfnamefont {J.~A.}\ \bibnamefont
  {Bianchini}}, \bibinfo {author} {\bibfnamefont {D.~J.}\ \bibnamefont
  {Whitney}}, \bibinfo {author} {\bibfnamefont {T.~D.}\ \bibnamefont
  {Breton}},\ and\ \bibinfo {author} {\bibfnamefont {B.~A.}\ \bibnamefont
  {Hilton-Brown}},\ }\bibfield  {title} {\bibinfo {title} {Toward inclusive
  science education: University scientists' views of students, instructional
  practices, and the nature of science},\ }\href@noop {} {\bibfield  {journal}
  {\bibinfo  {journal} {Science Education}\ }\textbf {\bibinfo {volume} {86}},\
  \bibinfo {pages} {42} (\bibinfo {year} {2002})}\BibitemShut {NoStop}%
\bibitem [{\citenamefont {Britner}\ and\ \citenamefont
  {Pajares}(2006)}]{britner2006}%
  \BibitemOpen
  \bibfield  {author} {\bibinfo {author} {\bibfnamefont {S.~L.}\ \bibnamefont
  {Britner}}\ and\ \bibinfo {author} {\bibfnamefont {F.}~\bibnamefont
  {Pajares}},\ }\bibfield  {title} {\bibinfo {title} {Sources of science
  self-efficacy beliefs of middle school students},\ }\href@noop {} {\bibfield
  {journal} {\bibinfo  {journal} {Journal of Research in Science Teaching: The
  Official Journal of the National Association for Research in Science
  Teaching}\ }\textbf {\bibinfo {volume} {43}},\ \bibinfo {pages} {485}
  (\bibinfo {year} {2006})}\BibitemShut {NoStop}%
\bibitem [{\citenamefont {Hilts}\ \emph {et~al.}(2018)\citenamefont {Hilts},
  \citenamefont {Part},\ and\ \citenamefont {Bernacki}}]{hilts2018}%
  \BibitemOpen
  \bibfield  {author} {\bibinfo {author} {\bibfnamefont {A.}~\bibnamefont
  {Hilts}}, \bibinfo {author} {\bibfnamefont {R.}~\bibnamefont {Part}},\ and\
  \bibinfo {author} {\bibfnamefont {M.~L.}\ \bibnamefont {Bernacki}},\
  }\bibfield  {title} {\bibinfo {title} {The roles of social influences on
  student competence, relatedness, achievement, and retention in {STEM}},\
  }\href@noop {} {\bibfield  {journal} {\bibinfo  {journal} {Science
  Education}\ }\textbf {\bibinfo {volume} {102}},\ \bibinfo {pages} {744}
  (\bibinfo {year} {2018})}\BibitemShut {NoStop}%
\bibitem [{\citenamefont {{Indiana University Center for Postsecondary
  Research}}(2018)}]{carnegie}%
  \BibitemOpen
  \bibfield  {author} {\bibinfo {author} {\bibnamefont {{Indiana University
  Center for Postsecondary Research}}},\ }\href
  {http://carnegieclassifications.iu.edu/} {\emph {\bibinfo {title} {The
  {C}arnegie {C}lassification of {I}nstitutions of {H}igher {E}ducation}}},\
  \bibinfo {type} {Tech. Rep.}\ (\bibinfo  {institution} {{Indiana University
  Center for Postsecondary Research}},\ \bibinfo {address} {Bloomington, IN},\
  \bibinfo {year} {2018})\BibitemShut {NoStop}%
\bibitem [{\citenamefont {Freedman}\ \emph {et~al.}(2007)\citenamefont
  {Freedman}, \citenamefont {Pisani},\ and\ \citenamefont
  {Purves}}]{freedman2007}%
  \BibitemOpen
  \bibfield  {author} {\bibinfo {author} {\bibfnamefont {D.}~\bibnamefont
  {Freedman}}, \bibinfo {author} {\bibfnamefont {R.}~\bibnamefont {Pisani}},\
  and\ \bibinfo {author} {\bibfnamefont {R.}~\bibnamefont {Purves}},\
  }\href@noop {} {\emph {\bibinfo {title} {Statistics}}},\ \bibinfo {edition}
  {4th}\ ed.\ (\bibinfo  {publisher} {W. W. Norton \& Co.},\ \bibinfo {year}
  {2007})\BibitemShut {NoStop}%
\bibitem [{\citenamefont {{R Core Team}}(2019)}]{rcran}%
  \BibitemOpen
  \bibfield  {author} {\bibinfo {author} {\bibnamefont {{R Core Team}}},\
  }\href {https://www.R-project.org/} {\emph {\bibinfo {title} {R: {A} Language
  and Environment for Statistical Computing}}},\ \bibinfo {organization} {R
  Foundation for Statistical Computing},\ \bibinfo {address} {Vienna, Austria}
  (\bibinfo {year} {2019})\BibitemShut {NoStop}%
\bibitem [{\citenamefont {Wickham}(2017)}]{tidyverse}%
  \BibitemOpen
  \bibfield  {author} {\bibinfo {author} {\bibfnamefont {H.}~\bibnamefont
  {Wickham}},\ }\href {https://CRAN.R-project.org/package=tidyverse} {\emph
  {\bibinfo {title} {{tidyverse}: {E}asily Install and Load the `tidyverse'}}}
  (\bibinfo {year} {2017}),\ \bibinfo {note} {{R} package version
  1.2.1}\BibitemShut {NoStop}%
\bibitem [{\citenamefont {Cohen}(1988)}]{cohen1988}%
  \BibitemOpen
  \bibfield  {author} {\bibinfo {author} {\bibfnamefont {J.}~\bibnamefont
  {Cohen}},\ }\href@noop {} {\emph {\bibinfo {title} {Statistical Power
  Analysis for the Behavioral Sciences}}},\ \bibinfo {edition} {2nd}\ ed.\
  (\bibinfo  {publisher} {Lawrence Erlbaum Associates},\ \bibinfo {year}
  {1988})\BibitemShut {NoStop}%
\bibitem [{\citenamefont {Marshman}\ \emph {et~al.}(2017)\citenamefont
  {Marshman}, \citenamefont {Kalender}, \citenamefont {Schunn}, \citenamefont
  {Nokes-Malach},\ and\ \citenamefont {Singh}}]{marshman2017}%
  \BibitemOpen
  \bibfield  {author} {\bibinfo {author} {\bibfnamefont {E.}~\bibnamefont
  {Marshman}}, \bibinfo {author} {\bibfnamefont {Z.~Y.}\ \bibnamefont
  {Kalender}}, \bibinfo {author} {\bibfnamefont {C.}~\bibnamefont {Schunn}},
  \bibinfo {author} {\bibfnamefont {T.}~\bibnamefont {Nokes-Malach}},\ and\
  \bibinfo {author} {\bibfnamefont {C.}~\bibnamefont {Singh}},\ }\bibfield
  {title} {\bibinfo {title} {A longitudinal analysis of students' motivational
  characteristics in introductory physics courses: Gender differences},\
  }\href@noop {} {\bibfield  {journal} {\bibinfo  {journal} {Canadian Journal
  of Physics}\ }\textbf {\bibinfo {volume} {96}},\ \bibinfo {pages} {391}
  (\bibinfo {year} {2017})}\BibitemShut {NoStop}%
\bibitem [{\citenamefont {Marshman}\ \emph {et~al.}(2018)\citenamefont
  {Marshman}, \citenamefont {Kalender}, \citenamefont {Nokes-Malach},
  \citenamefont {Schunn},\ and\ \citenamefont {Singh}}]{marshman2018}%
  \BibitemOpen
  \bibfield  {author} {\bibinfo {author} {\bibfnamefont {E.~M.}\ \bibnamefont
  {Marshman}}, \bibinfo {author} {\bibfnamefont {Z.~Y.}\ \bibnamefont
  {Kalender}}, \bibinfo {author} {\bibfnamefont {T.}~\bibnamefont
  {Nokes-Malach}}, \bibinfo {author} {\bibfnamefont {C.}~\bibnamefont
  {Schunn}},\ and\ \bibinfo {author} {\bibfnamefont {C.}~\bibnamefont
  {Singh}},\ }\bibfield  {title} {\bibinfo {title} {Female students with {A}'s
  have similar physics self-efficacy as male students with {C}'s in
  introductory courses: A cause for alarm?},\ }\href
  {https://doi.org/10.1103/PhysRevPhysEducRes.14.020123} {\bibfield  {journal}
  {\bibinfo  {journal} {Phys. Rev. Phys. Educ. Res.}\ }\textbf {\bibinfo
  {volume} {14}},\ \bibinfo {pages} {020123} (\bibinfo {year}
  {2018})}\BibitemShut {NoStop}%
\end{thebibliography}%

\clearpage
\onecolumngrid
\appendix

\section*[Appendix A]{Appendix A: Number of Current Majors and those who Added or Dropped each Major by Term}

\begin{table*}[b]

	\centering
	
	\vspace{-20mm} %to force the tables to occupy the same page as the appendix header
	
	\begin{minipage}{0.45\linewidth}%
		\begin{center}
	(a) \textbf{Chemistry}, $N_{\text{unique}} = 1090$
\end{center}

\begin{tabular}{c | r r r}
	\multirow{2}{*}{Term}	& \multicolumn{3}{c}{Number of Majors [\% of $N_{\text{unique}}$]} \\
	& Current	& Added	& Dropped	\\
	\hline
1	& 15 \hphantom{1}[1.4]	& 15 \hphantom{1}[1.4]	& 1 \hphantom{1}[0.1]	\\
2	& 102 \hphantom{1}[9.4]	& 87 \hphantom{1}[8.0]	& 1 \hphantom{1}[0.1]	\\
3	& 821 [75.3]	& 724 [66.4]	& 10 \hphantom{1}[0.9]	\\
4	& 912 [83.7]	& 150 [13.8]	& 61 \hphantom{1}[5.6]	\\
5	& 942 [86.4]	& 71 \hphantom{1}[6.5]	& 42 \hphantom{1}[3.9]	\\
6	& 926 [85.0]	& 24 \hphantom{1}[2.2]	& 41 \hphantom{1}[3.8]	\\
7	& 910 [83.5]	& 14 \hphantom{1}[1.3]	& 25 \hphantom{1}[2.3]	\\
8	& 884 [81.1]	& 1 \hphantom{1}[0.1]	& 15 \hphantom{1}[1.4]	\\
9	& 414 [38.0]	& 6 \hphantom{1}[0.6]	& 10 \hphantom{1}[0.9]	\\
10	& 161 [14.8]	& 1 \hphantom{1}[0.1]	& 7 \hphantom{1}[0.6]	\\
11	& 32 \hphantom{1}[2.9]	& 0 \hphantom{1}[0.0]	& 5 \hphantom{1}[0.5]	\\
12	& 16 \hphantom{1}[1.5]	& 1 \hphantom{1}[0.1]	& 4 \hphantom{1}[0.4]	\\
\end{tabular}
		
		\vspace{2mm}
		
		\begin{center}
	(c) \textbf{Engineering}, $N_{\text{unique}} = 3575$
\end{center}

\begin{tabular}{c | r r r}
	\multirow{2}{*}{Term}	& \multicolumn{3}{c}{Number of Majors [\% of $N_{\text{unique}}$]} \\
	& Current	& Added	& Dropped	\\
	\hline
1	& 3228 [90.3]	& 3228 [90.3]	& 47 \hphantom{1}[1.3]	\\
2	& 2980 [83.4]	& 14 \hphantom{1}[0.4]	& 262 \hphantom{1}[7.3]	\\
3	& 2968 [83.0]	& 159 \hphantom{1}[4.4]	& 171 \hphantom{1}[4.8]	\\
4	& 2902 [81.2]	& 55 \hphantom{1}[1.5]	& 121 \hphantom{1}[3.4]	\\
5	& 2902 [81.2]	& 64 \hphantom{1}[1.8]	& 64 \hphantom{1}[1.8]	\\
6	& 2901 [81.1]	& 27 \hphantom{1}[0.8]	& 28 \hphantom{1}[0.8]	\\
7	& 2881 [80.6]	& 21 \hphantom{1}[0.6]	& 40 \hphantom{1}[1.1]	\\
8	& 2847 [79.6]	& 5 \hphantom{1}[0.1]	& 25 \hphantom{1}[0.7]	\\
9	& 1609 [45.0]	& 4 \hphantom{1}[0.1]	& 15 \hphantom{1}[0.4]	\\
10	& 636 [17.8]	& 1 \hphantom{1}[0.0]	& 15 \hphantom{1}[0.4]	\\
11	& 114 \hphantom{1}[3.2]	& 0 \hphantom{1}[0.0]	& 8 \hphantom{1}[0.2]	\\
12	& 49 \hphantom{1}[1.4]	& 0 \hphantom{1}[0.0]	& 7 \hphantom{1}[0.2]	\\
\end{tabular}
	\end{minipage}\hfill%
	\begin{minipage}{0.45\linewidth}%
		\begin{center}
	(b) \textbf{Computer Science}, $N_{\text{unique}} = 621$
\end{center}

\begin{tabular}{c | r r r}
	\multirow{2}{*}{Term}	& \multicolumn{3}{c}{Number of Majors [\% of $N_{\text{unique}}$]} \\
	& Current	& Added	& Dropped	\\
	\hline
1	& 1 \hphantom{1}[0.2]	& 1 \hphantom{1}[0.2]	& 0 \hphantom{1}[0.0]	\\
2	& 20 \hphantom{1}[3.2]	& 19 \hphantom{1}[3.1]	& 0 \hphantom{1}[0.0]	\\
3	& 100 [16.1]	& 81 [13.0]	& 1 \hphantom{1}[0.2]	\\
4	& 214 [34.5]	& 116 [18.7]	& 5 \hphantom{1}[0.8]	\\
5	& 395 [63.6]	& 188 [30.3]	& 8 \hphantom{1}[1.3]	\\
6	& 459 [73.9]	& 74 [11.9]	& 10 \hphantom{1}[1.6]	\\
7	& 502 [80.8]	& 65 [10.5]	& 17 \hphantom{1}[2.7]	\\
8	& 495 [79.7]	& 27 \hphantom{1}[4.3]	& 12 \hphantom{1}[1.9]	\\
9	& 241 [38.8]	& 37 \hphantom{1}[6.0]	& 10 \hphantom{1}[1.6]	\\
10	& 147 [23.7]	& 11 \hphantom{1}[1.8]	& 11 \hphantom{1}[1.8]	\\
11	& 69 [11.1]	& 2 \hphantom{1}[0.3]	& 10 \hphantom{1}[1.6]	\\
12	& 36 \hphantom{1}[5.8]	& 2 \hphantom{1}[0.3]	& 2 \hphantom{1}[0.3]	\\
\end{tabular}
		
		\vspace{2mm}
		
		\begin{center}
	(d) \textbf{Mathematics}, $N_{\text{unique}} = 373$
\end{center}

\begin{tabular}{c | r r r}
	\multirow{2}{*}{Term}	& \multicolumn{3}{c}{Number of Majors [\% of $N_{\text{unique}}$]} \\
	& Current	& Added	& Dropped	\\
	\hline
1	& 24 \hphantom{1}[6.4]	& 24 \hphantom{1}[6.4]	& 0 \hphantom{1}[0.0]	\\
2	& 91 [24.4]	& 68 [18.2]	& 1 \hphantom{1}[0.3]	\\
3	& 185 [49.6]	& 102 [27.3]	& 10 \hphantom{1}[2.7]	\\
4	& 259 [69.4]	& 95 [25.5]	& 23 \hphantom{1}[6.2]	\\
5	& 270 [72.4]	& 36 \hphantom{1}[9.7]	& 26 \hphantom{1}[7]	\\
6	& 286 [76.7]	& 27 \hphantom{1}[7.2]	& 13 \hphantom{1}[3.5]	\\
7	& 278 [74.5]	& 14 \hphantom{1}[3.8]	& 15 \hphantom{1}[4.0]	\\
8	& 264 [70.8]	& 4 \hphantom{1}[1.1]	& 13 \hphantom{1}[3.5]	\\
9	& 81 [21.7]	& 8 \hphantom{1}[2.1]	& 9 \hphantom{1}[2.4]	\\
10	& 61 [16.4]	& 1 \hphantom{1}[0.3]	& 2 \hphantom{1}[0.5]	\\
11	& 18 \hphantom{1}[4.8]	& 1 \hphantom{1}[0.3]	& 6 \hphantom{1}[1.6]	\\
12	& 7 \hphantom{1}[1.9]	& 0 \hphantom{1}[0.0]	& 4 \hphantom{1}[1.1]	\\
\end{tabular}
	\end{minipage}
	
		\vspace{2mm}
		
		\begin{center}
	(e) \textbf{Physics \& Astronomy}, $N_{\text{unique}} = 186$
\end{center}

\begin{tabular}{c | r r r}
	\multirow{2}{*}{Term}	& \multicolumn{3}{c}{Number of Majors [\% of $N_{\text{unique}}$]} \\
	& Current	& Added	& Dropped	\\
	\hline
1	& 17 \hphantom{1}[9.1]	& 17 \hphantom{1}[9.1]	& 0 \hphantom{1}[0.0]	\\
2	& 74 [39.8]	& 58 [31.2]	& 1 \hphantom{1}[0.5]	\\
3	& 124 [66.7]	& 62 [33.3]	& 14 \hphantom{1}[7.5]	\\
4	& 151 [81.2]	& 34 [18.3]	& 8 \hphantom{1}[4.3]	\\
5	& 148 [79.6]	& 8 \hphantom{1}[4.3]	& 12 \hphantom{1}[6.5]	\\
6	& 144 [77.4]	& 4 \hphantom{1}[2.2]	& 8 \hphantom{1}[4.3]	\\
7	& 134 [72.0]	& 2 \hphantom{1}[1.1]	& 12 \hphantom{1}[6.5]	\\
8	& 128 [68.8]	& 1 \hphantom{1}[0.5]	& 6 \hphantom{1}[3.2]	\\
9	& 46 [24.7]	& 1 \hphantom{1}[0.5]	& 4 \hphantom{1}[2.2]	\\
10	& 38 [20.4]	& 0 \hphantom{1}[0.0]	& 2 \hphantom{1}[1.1]	\\
11	& 3 \hphantom{1}[1.6]	& 0 \hphantom{1}[0.0]	& 1 \hphantom{1}[0.5]	\\
12	& 1 \hphantom{1}[0.5]	& 0 \hphantom{1}[0.0]	& 2 \hphantom{1}[1.1]	\\
\end{tabular}
	
	\caption{\label{table_adddrop_1}
		For each term from 1 to 12, the current number of declared majors (``Current'') is shown along with the number of majors who newly declared in that term (``Added'') and the number of former majors who dropped the major as of that term (``Dropped'').
		In square brackets next to each measure is the percentage of all unique students who declared that major.
		The five sub-tables show this information for five different majors: (a) chemistry, (b) computer science, (c) engineering, (d) mathematics, and (e) physics and astronomy.
		For example, in (c) we can see that in term 2, there were 2980 students with a declared engineering major, which represents 83.4\% of all students who ever declared engineering.
		Further, 14 students (0.4\% of all engineering students) who did not declare in term 1 added the major in term 2, and 262 students (7.3\% of all engineering students) who were engineering majors in term 1 dropped the major in term 2.
		}
\end{table*}

\begin{table*}[b]
	
	\centering

	\begin{minipage}{0.45\linewidth}
		\begin{center}
	(a) \textbf{Biological Sciences}, $N_{\text{unique}} = 2249$
\end{center}

\begin{tabular}{c | r r r}
	\multirow{2}{*}{Term}	& \multicolumn{3}{c}{Number of Majors [\% of $N_{\text{unique}}$]} \\
	& Current	& Added	& Dropped	\\
	\hline
1	& 17 \hphantom{1}[0.8]	& 17 \hphantom{1}[0.8]	& 0 \hphantom{1}[0.0]	\\
2	& 69 \hphantom{1}[3.1]	& 54 \hphantom{1}[2.4]	& 3 \hphantom{1}[0.1]	\\
3	& 1224 [54.4]	& 1159 [51.5]	& 12 \hphantom{1}[0.5]	\\
4	& 1730 [76.9]	& 582 [25.9]	& 88 \hphantom{1}[3.9]	\\
5	& 1968 [87.5]	& 311 [13.8]	& 84 \hphantom{1}[3.7]	\\
6	& 1960 [87.1]	& 71 \hphantom{1}[3.2]	& 85 \hphantom{1}[3.8]	\\
7	& 1910 [84.9]	& 44 \hphantom{1}[2.0]	& 76 \hphantom{1}[3.4]	\\
8	& 1824 [81.1]	& 11 \hphantom{1}[0.5]	& 49 \hphantom{1}[2.2]	\\
9	& 354 [15.7]	& 7 \hphantom{1}[0.3]	& 34 \hphantom{1}[1.5]	\\
10	& 225 [10.0]	& 2 \hphantom{1}[0.1]	& 16 \hphantom{1}[0.7]	\\
11	& 41 \hphantom{1}[1.8]	& 1 \hphantom{1}[0.0]	& 6 \hphantom{1}[0.3]	\\
12	& 13 \hphantom{1}[0.6]	& 0 \hphantom{1}[0.0]	& 5 \hphantom{1}[0.2]	\\
\end{tabular}
		
		\vspace{3mm}
		
		\begin{center}
	(c) \textbf{Psychology}, $N_{\text{unique}} = 1784$
\end{center}

\begin{tabular}{c | r r r}
	\multirow{2}{*}{Term}	& \multicolumn{3}{c}{Number of Majors [\% of $N_{\text{unique}}$]} \\
	& Current	& Added	& Dropped	\\
	\hline
1	& 6 \hphantom{1}[0.3]	& 6 \hphantom{1}[0.3]	& 0 \hphantom{1}[0.0]	\\
2	& 62 \hphantom{1}[3.5]	& 56 \hphantom{1}[3.1]	& 0 \hphantom{1}[0.0]	\\
3	& 404 [22.6]	& 343 [19.2]	& 4 \hphantom{1}[0.2]	\\
4	& 1011 [56.7]	& 620 [34.8]	& 25 \hphantom{1}[1.4]	\\
5	& 1446 [81.1]	& 458 [25.7]	& 23 \hphantom{1}[1.3]	\\
6	& 1594 [89.3]	& 174 \hphantom{1}[9.8]	& 29 \hphantom{1}[1.6]	\\
7	& 1657 [92.9]	& 95 \hphantom{1}[5.3]	& 13 \hphantom{1}[0.7]	\\
8	& 1552 [87.0]	& 23 \hphantom{1}[1.3]	& 28 \hphantom{1}[1.6]	\\
9	& 265 [14.9]	& 18 \hphantom{1}[1.0]	& 12 \hphantom{1}[0.7]	\\
10	& 138 \hphantom{1}[7.7]	& 4 \hphantom{1}[0.2]	& 10 \hphantom{1}[0.6]	\\
11	& 32 \hphantom{1}[1.8]	& 0 \hphantom{1}[0.0]	& 6 \hphantom{1}[0.3]	\\
12	& 16 \hphantom{1}[0.9]	& 0 \hphantom{1}[0.0]	& 5 \hphantom{1}[0.3]	\\
\end{tabular}
	\end{minipage}\hfill%
	\begin{minipage}{0.45\linewidth}%
		\begin{center}
	(b) \textbf{Economics}, $N_{\text{unique}} = 965$
\end{center}

\begin{tabular}{c | r r r}
	\multirow{2}{*}{Term}	& \multicolumn{3}{c}{Number of Majors [\% of $N_{\text{unique}}$]} \\
	& Current	& Added	& Dropped	\\
	\hline
1	& 17 \hphantom{1}[1.8]	& 17 \hphantom{1}[1.8]	& 0 \hphantom{1}[0.0]	\\
2	& 104 [10.8]	& 89 \hphantom{1}[9.2]	& 5 \hphantom{1}[0.5]	\\
3	& 306 [31.7]	& 210 [21.8]	& 9 \hphantom{1}[0.9]	\\
4	& 562 [58.2]	& 271 [28.1]	& 22 \hphantom{1}[2.3]	\\
5	& 708 [73.4]	& 175 [18.1]	& 31 \hphantom{1}[3.2]	\\
6	& 766 [79.4]	& 89 \hphantom{1}[9.2]	& 33 \hphantom{1}[3.4]	\\
7	& 766 [79.4]	& 70 \hphantom{1}[7.3]	& 53 \hphantom{1}[5.5]	\\
8	& 717 [74.3]	& 28 \hphantom{1}[2.9]	& 30 \hphantom{1}[3.1]	\\
9	& 183 [19.0]	& 12 \hphantom{1}[1.2]	& 22 \hphantom{1}[2.3]	\\
10	& 92 \hphantom{1}[9.5]	& 4 \hphantom{1}[0.4]	& 15 \hphantom{1}[1.6]	\\
11	& 34 \hphantom{1}[3.5]	& 4 \hphantom{1}[0.4]	& 4 \hphantom{1}[0.4]	\\
12	& 11 \hphantom{1}[1.1]	& 0 \hphantom{1}[0.0]	& 9 \hphantom{1}[0.9]	\\
\end{tabular}
		
		\vspace{3mm}
		
		\begin{center}
	(d) \textbf{Other STEM}, $N_{\text{unique}} = 1543$
\end{center}

\begin{tabular}{c | r r r}
	\multirow{2}{*}{Term}	& \multicolumn{3}{c}{Number of Majors [\% of $N_{\text{unique}}$]} \\
	& Current	& Added	& Dropped	\\
	\hline
1	& 16 \hphantom{1}[1.0]	& 16 \hphantom{1}[1.0]	& 0 \hphantom{1}[0.0]	\\
2	& 143 \hphantom{1}[9.3]	& 128 \hphantom{1}[8.3]	& 4 \hphantom{1}[0.3]	\\
3	& 781 [50.6]	& 648 [42.0]	& 13 \hphantom{1}[0.8]	\\
4	& 1212 [78.5]	& 476 [30.8]	& 52 \hphantom{1}[3.4]	\\
5	& 1345 [87.2]	& 188 [12.2]	& 58 \hphantom{1}[3.8]	\\
6	& 1333 [86.4]	& 42 \hphantom{1}[2.7]	& 56 \hphantom{1}[3.6]	\\
7	& 1284 [83.2]	& 27 \hphantom{1}[1.7]	& 56 \hphantom{1}[3.6]	\\
8	& 1222 [79.2]	& 9 \hphantom{1}[0.6]	& 23 \hphantom{1}[1.5]	\\
9	& 263 [17.0]	& 9 \hphantom{1}[0.6]	& 19 \hphantom{1}[1.2]	\\
10	& 135 \hphantom{1}[8.7]	& 0 \hphantom{1}[0.0]	& 14 \hphantom{1}[0.9]	\\
11	& 36 \hphantom{1}[2.3]	& 1 \hphantom{1}[0.1]	& 6 \hphantom{1}[0.4]	\\
12	& 12 \hphantom{1}[0.8]	& 0 \hphantom{1}[0.0]	& 11 \hphantom{1}[0.7]	\\
\end{tabular}
	\end{minipage}
	
		\vspace{3mm}
		
		\begin{center}
	(e) \textbf{Non-STEM}, $N_{\text{unique}} = 5197$
\end{center}

\begin{tabular}{c | r r r}
	\multirow{2}{*}{Term}	& \multicolumn{3}{c}{Number of Majors [\% of $N_{\text{unique}}$]} \\
	& Current	& Added	& Dropped	\\
	\hline
1	& 259 \hphantom{1}[5.0]	& 259 \hphantom{1}[5.0]	& 1 \hphantom{1}[0.0]	\\
2	& 1038 [20.0]	& 811 [15.6]	& 41 \hphantom{1}[0.8]	\\
3	& 2502 [48.1]	& 1507 [29]	& 51 \hphantom{1}[1.0]	\\
4	& 3600 [69.3]	& 1186 [22.8]	& 110 \hphantom{1}[2.1]	\\
5	& 4225 [81.3]	& 759 [14.6]	& 84 \hphantom{1}[1.6]	\\
6	& 4437 [85.4]	& 308 \hphantom{1}[5.9]	& 70 \hphantom{1}[1.3]	\\
7	& 4449 [85.6]	& 221 \hphantom{1}[4.3]	& 113 \hphantom{1}[2.2]	\\
8	& 4201 [80.8]	& 109 \hphantom{1}[2.1]	& 110 \hphantom{1}[2.1]	\\
9	& 818 [15.7]	& 45 \hphantom{1}[0.9]	& 68 \hphantom{1}[1.3]	\\
10	& 454 \hphantom{1}[8.7]	& 11 \hphantom{1}[0.2]	& 32 \hphantom{1}[0.6]	\\
11	& 115 \hphantom{1}[2.2]	& 5 \hphantom{1}[0.1]	& 15 \hphantom{1}[0.3]	\\
12	& 58 \hphantom{1}[1.1]	& 3 \hphantom{1}[0.1]	& 12 \hphantom{1}[0.2]	\\
\end{tabular}
	
	\caption{\label{table_adddrop_2}
		For each term from 1 to 12, the current number of declared majors (``Current'') is shown along with the number of majors who newly declared in that term (``Added'') and the number of former majors who dropped the major as of that term (``Dropped'').
		In square brackets next to each measure is the percentage of all unique students who declared that major (or cluster of majors as in (d) and (e)).
		The five sub-tables show this information for five different majors or clusters of majors: (a) biological sciences, (b) economics, (c) psychology, (d) other STEM disciplines not listed separately in Tables III and IV, and (e) other non-STEM majors excluding psychology.
		}
\end{table*}

\clearpage

\section*[Appendix B]{Appendix B: Summary Tables Pertaining to different Declared Majors}

\begin{table*}[b]

		\begin{tabular}{l | r r r r}
	\textbf{All Students}	& Unique	& Peak Concurrent	& Peak Added		& Peak Dropped	\\
	Major					& Majors 	& Majors [Term]		& Majors [Term]	& Majors [Term]	\\
	\hline
Biological Sciences	& 2249	& 1968 [5]	& 1159 [3]	& 88 [4]	\\
Chemistry	& 1090	& 942 [5]	& 724 [3]	& 61 [4]	\\
Computer Science	& 621	& 502 [7]	& 188 [5]	& 17 [7]	\\
Economics	& 965	& 766 [7]	& 271 [4]	& 53 [7]	\\
Engineering	& 3575	& 3228 [1]	& 3228 [1]	& 262 [2]	\\
Mathematics	& 373	& 286 [6]	& 102 [3]	& 26 [5]	\\
Physics and Astronomy	& 186	& 151 [4]	& 62 [3]	& 14 [3]	\\
Psychology	& 1784	& 1657 [7]	& 620 [4]	& 29 [6]	\\
Other STEM	& 1543	& 1345 [5]	& 648 [3]	& 58 [5]	\\
Non-STEM	& 5197	& 4449 [7]	& 1507 [3]	& 113 [7]	\\
\end{tabular}
		
	\caption{\label{table_appendix_all_main}
	Summary counts for all students.
	For each major, the total number of unique students is listed along with peak concurrent majors, added majors, and dropped majors, as well as the term in which the peak occurs in brackets.
	For example, in biological sciences, there were 2249 individual students in the sample who had ever declared the major.
	1968 of those students declared biological science majors in term 5 (peak term for concurrent majors), which is higher than the number of majors declared in any other term.
	Further, 1159 of those students added the major in term 3 (peak term for adding this major) and 88 of those students dropped the major in term 4 (peak term for dropping this major).
	}
\end{table*}

\begin{table*}[b]

		\begin{tabular}{l | r r r r}
	\textbf{Men Only}	& Unique	& Peak Concurrent	& Peak Added		& Peak Dropped	\\
	Major					& Majors 	& Majors [Term]		& Majors [Term]	& Majors [Term]	\\
	\hline
Biological Sciences	& 1115	& 962 [6]	& 620 [3]	& 52 [4]	\\
Chemistry	& 647	& 567 [6]	& 434 [3]	& 28 [4]	\\
Computer Science	& 512	& 413 [7]	& 156 [5]	& 14 [7]	\\
Economics	& 661	& 514 [7]	& 189 [4]	& 36 [7]	\\
Engineering	& 2678	& 2410 [1]	& 2410 [1]	& 189 [2]	\\
Mathematics	& 242	& 185 [6]	& 59 [4]	& 13 [5]	\\
Physics and Astronomy	& 150	& 124 [4]	& 54 [3]	& 11 [7]	\\
Psychology	& 490	& 447 [7]	& 141 [4]	& 14 [8]	\\
Other STEM	& 781	& 678 [5]	& 320 [3]	& 38 [7]	\\
Non-STEM	& 2235	& 1853 [7]	& 613 [3]	& 63 [7]	\\
\end{tabular}
				
	\caption{\label{table_appendix_M_main}
	Summary counts for men.
	For each major, the total number of unique male students is listed along with peak concurrent majors, added majors, and dropped majors, as well as the term in which the peak occurs in brackets.
	}
\end{table*}

\begin{table*}[b]

		\begin{tabular}{l | r r r r}
	\textbf{Women Only}	& Unique	& Peak Concurrent	& Peak Added		& Peak Dropped	\\
	Major					& Majors 	& Majors [Term]		& Majors [Term]	& Majors [Term]	\\
	\hline
Biological Sciences	& 1134	& 1013 [5]	& 539 [3]	& 53 [6]	\\
Chemistry	& 443	& 377 [5]	& 290 [3]	& 33 [4]	\\
Computer Science	& 109	& 89 [7]	& 32 [5]	& 3 [7]	\\
Economics	& 304	& 252 [7]	& 82 [4]	& 17 [7]	\\
Engineering	& 897	& 818 [1]	& 818 [1]	& 73 [2]	\\
Mathematics	& 131	& 101 [6]	& 47 [3]	& 13 [5]	\\
Physics and Astronomy	& 36	& 28 [6]	& 13 [2]	& 4 [3]	\\
Psychology	& 1294	& 1210 [7]	& 479 [4]	& 20 [6]	\\
Other STEM	& 762	& 667 [5]	& 328 [3]	& 34 [5]	\\
Non-STEM	& 2962	& 2596 [7]	& 894 [3]	& 56 [4]	\\
\end{tabular}
				
	\caption{\label{table_appendix_F_main}
	Summary counts for women.
	For each major, the total number of unique female students is listed along with peak concurrent majors, added majors, and dropped majors, as well as the term in which the peak occurs in brackets.
	}
\end{table*}

\clearpage

\section*[Appendix C]{Appendix C: Degrees Earned by Students who Dropped a Major}

\begin{table*}[b]

	\vspace{-5mm} %to force the tables to occupy the same page as the appendix header
	
		%\begin{tabular}{l r | r r r r r r r r r r r}
\begin{tabular}{l r | R{1cm} R{1cm} R{1cm} R{1cm} R{1cm} R{1cm} R{1cm} R{1cm} R{1cm} R{1cm} R{1cm}}
%	\textbf{All Students}		& 					& \multicolumn{11}{c}{\% of $N_{\text{drop}}$ That Earned Degree in Major} \\
	\textbf{All Students}		& 					& \multicolumn{11}{c}{\% of $N_{\text{drop}}$ in a Given Major That Subsequently Earned Degree in Each Major} \\
								&					& & & & & & & & & Other & Non-& No \\
	Major						& $N_{\text{drop}}$	& Bio 
													& Chem 
													& CS
													& Econ
													& Engr
													& Math
													& Phys
													& Psych
													& STEM
													& STEM
													& Degree
													\\
	\hline
Biological Sciences	& 458	& 2.6	& 3.9	& 3.1	& 1.5	& 10.0	& 2.2	& 0.2	& 7.9	& 8.1	& 44.1	& 27.1	\\
Chemistry	& 222	& 10.8	& 1.4	& 4.5	& 1.8	& 11.7	& 2.3	& 0.9	& 5.0	& 7.7	& 37.4	& 26.6	\\
Computer Science	& 86	& 5.8	& 0.0	& 1.2	& 2.3	& 2.3	& 3.5	& 1.2	& 2.3	& 5.8	& 26.7	& 53.5	\\
Economics	& 233	& 2.6	& 1.3	& 3.4	& 1.3	& 2.6	& 6.9	& 0.0	& 4.3	& 7.3	& 47.2	& 31.8	\\
Engineering	& 803	& 4.1	& 2.6	& 8.2	& 5.4	& 0.9	& 2.4	& 0.7	& 2.5	& 4.9	& 29.6	& 46.3	\\
Mathematics	& 122	& 10.7	& 5.7	& 4.9	& 15.6	& 5.7	& 4.1	& 1.6	& 7.4	& 19.7	& 20.5	& 25.4	\\
Physics and Astronomy	& 70	& 1.4	& 4.3	& 4.3	& 4.3	& 11.4	& 15.7	& 1.4	& 2.9	& 2.9	& 24.3	& 37.1	\\
Psychology	& 155	& 1.9	& 0.0	& 3.2	& 1.3	& 0.6	& 0.0	& 0.6	& 9.7	& 2.6	& 35.5	& 51.6	\\
Other STEM	& 312	& 11.5	& 1.3	& 3.2	& 11.2	& 2.6	& 5.4	& 0.0	& 15.1	& 1.0	& 36.2	& 26.6	\\
Non-STEM	& 707	& 6.6	& 1.3	& 4.4	& 4.7	& 2.7	& 1.6	& 1.0	& 12.2	& 5.1	& 8.8	& 59.3	\\
\end{tabular}
	
	\caption{\label{table_appendix_all_supp}
	Trajectory of all students who dropped a major.
	For each major, the total number of students in the dataset who dropped that major ($N_{\text{drop}}$) is listed along with the percentage of $N_{\text{drop}}$ who ultimately earned a degree in each major or earned no degree.
	For example, there were 458 students who ever dropped their major in biological sciences.
	Of those 458 students, 2.6\% went on to earn a degree in biological sciences (i.e., they later declared that major again after dropping it at an earlier point).
	Similarly, 3.9\% of them earned a degree in chemistry, 3.1\% in computer science, 1.5\% in economics, 10.0\% in engineering, and so forth.
	Finally, 27.1\% of those 458 students that dropped a biological sciences major ultimately did not earn a degree from the university.
	}
\end{table*}

\begin{table*}[b]
		%\begin{tabular}{l r | r r r r r r r r r r r}
\begin{tabular}{l r | R{1cm} R{1cm} R{1cm} R{1cm} R{1cm} R{1cm} R{1cm} R{1cm} R{1cm} R{1cm} R{1cm}}
%	\textbf{Men Only}		& 					& \multicolumn{11}{c}{\% of $N_{\text{drop}}$ That Earned Degree in Major} \\
	\textbf{Men Only}		& 					& \multicolumn{11}{c}{\% of $N_{\text{drop}}$ in a Given Major That Subsequently Earned Degree in Each Major} \\
								&					& & & & & & & & & Other & Non-& No \\
	Major						& $N_{\text{drop}}$	& Bio 
													& Chem 
													& CS
													& Econ
													& Engr
													& Math
													& Phys
													& Psych
													& STEM
													& STEM
													& Degree
													\\
	\hline
Biological Sciences	& 232	& 3.9	& 6.0	& 5.6	& 0.9	& 16.4	& 2.6	& 0.4	& 3.4	& 6.9	& 36.2	& 30.6	\\
Chemistry	& 118	& 8.5	& 2.5	& 8.5	& 2.5	& 11.0	& 2.5	& 0.8	& 4.2	& 7.6	& 31.4	& 28.8	\\
Computer Science	& 72	& 6.9	& 0.0	& 1.4	& 2.8	& 2.8	& 2.8	& 1.4	& 0.0	& 4.2	& 27.8	& 55.6	\\
Economics	& 177	& 2.3	& 0.6	& 4.5	& 1.7	& 2.8	& 6.8	& 0.0	& 3.4	& 6.2	& 45.8	& 33.9	\\
Engineering	& 609	& 2.8	& 2.0	& 9.7	& 5.9	& 0.5	& 2.3	& 0.8	& 1.3	& 3.9	& 28.6	& 49.3	\\
Mathematics	& 80	& 10.0	& 7.5	& 5.0	& 11.2	& 7.5	& 5.0	& 2.5	& 3.8	& 17.5	& 20.0	& 31.2	\\
Physics and Astronomy	& 59	& 1.7	& 5.1	& 5.1	& 5.1	& 11.9	& 15.3	& 1.7	& 1.7	& 3.4	& 22.0	& 37.3	\\
Psychology	& 61	& 0.0	& 0.0	& 8.2	& 3.3	& 0.0	& 0.0	& 1.6	& 13.1	& 0.0	& 34.4	& 47.5	\\
Other STEM	& 164	& 9.8	& 1.2	& 4.9	& 17.7	& 4.3	& 7.9	& 0.0	& 11.0	& 0.6	& 26.2	& 29.3	\\
Non-STEM	& 383	& 6.5	& 1.0	& 6.3	& 6.0	& 3.1	& 1.8	& 1.3	& 6.0	& 4.2	& 7.3	& 62.7	\\
\end{tabular}
	
	\caption{\label{table_appendix_M_supp}
	Trajectory of all men who dropped a major.
	For each major, the total number of male students in the dataset who dropped that major ($N_{\text{drop}}$) is listed along with the percentage of $N_{\text{drop}}$ who ultimately earned a degree in each major or earned no degree.
	}
\end{table*}

\begin{table*}[b]
		%\begin{tabular}{l r | r r r r r r r r r r r}
\begin{tabular}{l r | R{1cm} R{1cm} R{1cm} R{1cm} R{1cm} R{1cm} R{1cm} R{1cm} R{1cm} R{1cm} R{1cm}}
%	\textbf{Women Only}		& 					& \multicolumn{11}{c}{\% of $N_{\text{drop}}$ That Earned Degree in Major} \\
	\textbf{Women Only}		& 					& \multicolumn{11}{c}{\% of $N_{\text{drop}}$ in a Given Major That Subsequently Earned Degree in Each Major} \\
								&					& & & & & & & & & Other & Non-& No \\
	Major						& $N_{\text{drop}}$	& Bio 
													& Chem 
													& CS
													& Econ
													& Engr
													& Math
													& Phys
													& Psych
													& STEM
													& STEM
													& Degree
													\\
	\hline
Biological Sciences	& 226	& 1.3	& 1.8	& 0.4	& 2.2	& 3.5	& 1.8	& 0.0	& 12.4	& 9.3	& 52.2	& 23.5	\\
Chemistry	& 104	& 13.5	& 0.0	& 0.0	& 1.0	& 12.5	& 1.9	& 1.0	& 5.8	& 7.7	& 44.2	& 24.0	\\
Computer Science	& 14	& 0.0	& 0.0	& 0.0	& 0.0	& 0.0	& 7.1	& 0.0	& 14.3	& 14.3	& 21.4	& 42.9	\\
Economics	& 56	& 3.6	& 3.6	& 0.0	& 0.0	& 1.8	& 7.1	& 0.0	& 7.1	& 10.7	& 51.8	& 25.0	\\
Engineering	& 194	& 8.2	& 4.6	& 3.6	& 3.6	& 2.1	& 2.6	& 0.5	& 6.2	& 7.7	& 33.0	& 37.1	\\
Mathematics	& 42	& 11.9	& 2.4	& 4.8	& 23.8	& 2.4	& 2.4	& 0.0	& 14.3	& 23.8	& 21.4	& 14.3	\\
Physics and Astronomy	& 11	& 0.0	& 0.0	& 0.0	& 0.0	& 9.1	& 18.2	& 0.0	& 9.1	& 0.0	& 36.4	& 36.4	\\
Psychology	& 94	& 3.2	& 0.0	& 0.0	& 0.0	& 1.1	& 0.0	& 0.0	& 7.4	& 4.3	& 36.2	& 54.3	\\
Other STEM	& 148	& 13.5	& 1.4	& 1.4	& 4.1	& 0.7	& 2.7	& 0.0	& 19.6	& 1.4	& 47.3	& 23.6	\\
Non-STEM	& 324	& 6.8	& 1.5	& 2.2	& 3.1	& 2.2	& 1.2	& 0.6	& 19.4	& 6.2	& 10.5	& 55.2	\\
\end{tabular}
	
	\caption{\label{table_appendix_F_supp}
	Trajectory of all women who dropped a major.
	For each major, the total number of female students in the dataset who dropped that major ($N_{\text{drop}}$) is listed along with the percentage of $N_{\text{drop}}$ who ultimately earned a degree in each major or earned no degree.
	}
\end{table*}

\end{document}